\documentclass[aps,prl,twocolumn,preprintnumbers,amsmath,amssymb,
superscriptaddress,floatfix]{revtex4}%

\usepackage{graphicx}
\usepackage{dcolumn}
\usepackage{amsmath}
\usepackage{amssymb}
\usepackage{color}
\usepackage{multirow}
\usepackage{url}

\newcommand\hly{\bgroup\markoverwith
  {\textcolor{yellow}{\rule[-.5ex]{.1pt}{2.5ex}}}\ULon}

\newcommand{\mSR}{$\mu$SR}

\newcommand{\Tc}{$T_{\mathrm c}$}

\newcommand{\scg}{superconducting }
\newcommand{\BiSe}{Bi$_2$Se$_3$}

\newcommand{\refsubfig}[2]{\hyperref[#1]{\ref*{#1}#2}}

\usepackage[utf8]{inputenc}
\usepackage{siunitx} 
\usepackage{verbatim} 
\usepackage[normalem]{ulem}

\usepackage{placeins} 
\usepackage{float} 

\usepackage[hidelinks]{hyperref}
\hypersetup{
colorlinks=true,    
linkcolor=blue,
citecolor=blue,
urlcolor=blue,
pdfborder={0 0 0},
bookmarksopen=true,
bookmarksopenlevel=2,
pdfpagemode=UseOutlines,
pdfstartview={Fit},
pdftitle={Proximity-Induced Odd-Frequency Superconductivity in a Topological Insulator},
pdfauthor={J. A. Krieger},
pdfsubject={Proximity-Induced Odd-Frequency Superconductivity in a Topological Insulator},
pdfkeywords={Odd-Frequency Superconductivity, Low-Energy Muons, Topological 
Insulator, Superconducting Proximity},
pdfhighlight= /I
}

\begin{document}


\title {Proximity-Induced Odd-Frequency Superconductivity in a Topological Insulator}

\author{Jonas~A.~Krieger}
\affiliation{Laboratory for Muon Spin Spectroscopy, Paul Scherrer 
Institute, CH-5232 Villigen PSI, Switzerland}
\affiliation{Laboratorium f\"ur Festk\"orperphysik,  ETH Z\"urich, CH-8093
Z\"urich, Switzerland}
\affiliation{Swiss Light Source, Paul Scherrer Institute, 
CH-5232 Villigen PSI, Switzerland}
\author{Anna~Pertsova}
\affiliation{Nordita, Roslagstullsbacken 23, SE-106 91 Stockholm, Sweden}
\author{Sean~R.~Giblin}
\affiliation{School of Physics and Astronomy, Cardiff University, Cardiff 
CF24 3AA, United Kingdom}
\author{Max D\"obeli}
\affiliation{Ion Beam Physics, ETH Z\"urich, Otto-Stern-Weg 5, CH-8093 Z\"urich, Switzerland}
\author{Thomas~Prokscha}
\affiliation{Laboratory for Muon Spin Spectroscopy, Paul Scherrer 
Institute, CH-5232 Villigen PSI, CH}
\author{Christof W. Schneider}
\affiliation{Laboratory for Multiscale Materials Experiments, Paul Scherrer Institute, CH-5232 Villigen PSI, Switzerland}
\author{Andreas~Suter}
\affiliation{Laboratory for Muon Spin Spectroscopy, Paul Scherrer 
Institute, CH-5232 Villigen PSI, CH}
\author{Thorsten~Hesjedal}
\affiliation{Department of Physics, Clarendon Laboratory, University of 
Oxford, Oxford OX1 3PU, United Kingdom}
\author{Alexander~V.~Balatsky}
\email[Correspondnig author: ]{avb@nordita.org}
\affiliation{Nordita, Roslagstullsbacken 23, SE-106 91 Stockholm, Sweden}
\affiliation{Department of Physics, University of Connecticut, Storrs, Connecticut 06268, USA}
\author{Zaher~Salman}
\email[Correspondnig author: ]{zaher.salman@psi.ch}
\affiliation{Laboratory for Muon Spin Spectroscopy, Paul Scherrer 
Institute, CH-5232 Villigen PSI, CH}

\date{\today}

\begin{abstract}
At an interface between a topological insulator (TI) and a
conventional superconductor (SC), superconductivity has been predicted
to change dramatically and exhibit novel correlations.  In particular,
the induced superconductivity by an $s$-wave SC in a TI can develop an
order parameter with a $p$-wave component. Here we present
experimental evidence for an unexpected proximity-induced novel
superconducting state in a thin layer of the prototypical TI, \BiSe\,
proximity coupled to Nb. From depth-resolved magnetic field
measurements below the superconducting transition temperature of Nb,
we observe a local enhancement of the magnetic field in \BiSe\ that
exceeds the externally applied field, thus supporting the existence of
an intrinsic paramagnetic Meissner effect arising from an
odd-frequency superconducting state. Our experimental results are
complemented by theoretical calculations supporting the appearance of
such a component at the interface which extends into the TI. This
state is topologically distinct from the conventional
Bardeen–Cooper–Schrieffer state it originates from. To the best of our
knowledge, these findings present a first observation of bulk
odd-frequency superconductivity in a TI. We thus reaffirm the
potential of the TI-SC interface as a versatile platform to produce
novel superconducting states.
\end{abstract}

\maketitle

In a conventional superconductor (SC), the electronic excitations are
usually described as condensation of Cooper
pairs~\cite{Tinkham2004}. Fermi statistics imply symmetry
constraints on permutation properties of the pair wave function, thus
limiting the possible SC classes~\cite{Berezinskii1974}. According to
conventional classification, states with even parity ($s$-, $d$-wave)
must be in a spin-singlet configuration while states with odd parity
($p$-, $f$-wave) must be in a spin-triplet configuration.  However,
additional classes are possible when permutation with respect to time
and, if present, orbital degrees of freedom are included. This general
classification allows for the odd-frequency, or Berezinskii, state
characterized by superconducting pairing which is nonlocal and odd in
time~\cite{Berezinskii1974,Kirkpatrick1991PRL,Belitz1992PRB,Balatsky1992PRB,LinderBalatsky2019PRM}.
Odd-frequency pairing gives rise to SC states with symmetries
different from conventional states, for example triplet $s$-wave
\cite{Kirkpatrick1991PRL,Belitz1992PRB} and singlet $p$-wave
\cite{Balatsky1992PRB} states. The Berezinskii state is currently
recognized as an inherently dynamical order that can be realized in a
variety of systems, including bulk SCs, heterostructures, and
dynamically driven systems. It is especially relevant at interfaces,
where locally broken symmetries can influence the type of
pairing~\cite{Tanaka2007b,LinderBalatsky2019PRM}.

\begin{figure*}[htb]
  \includegraphics[width=.8\linewidth]{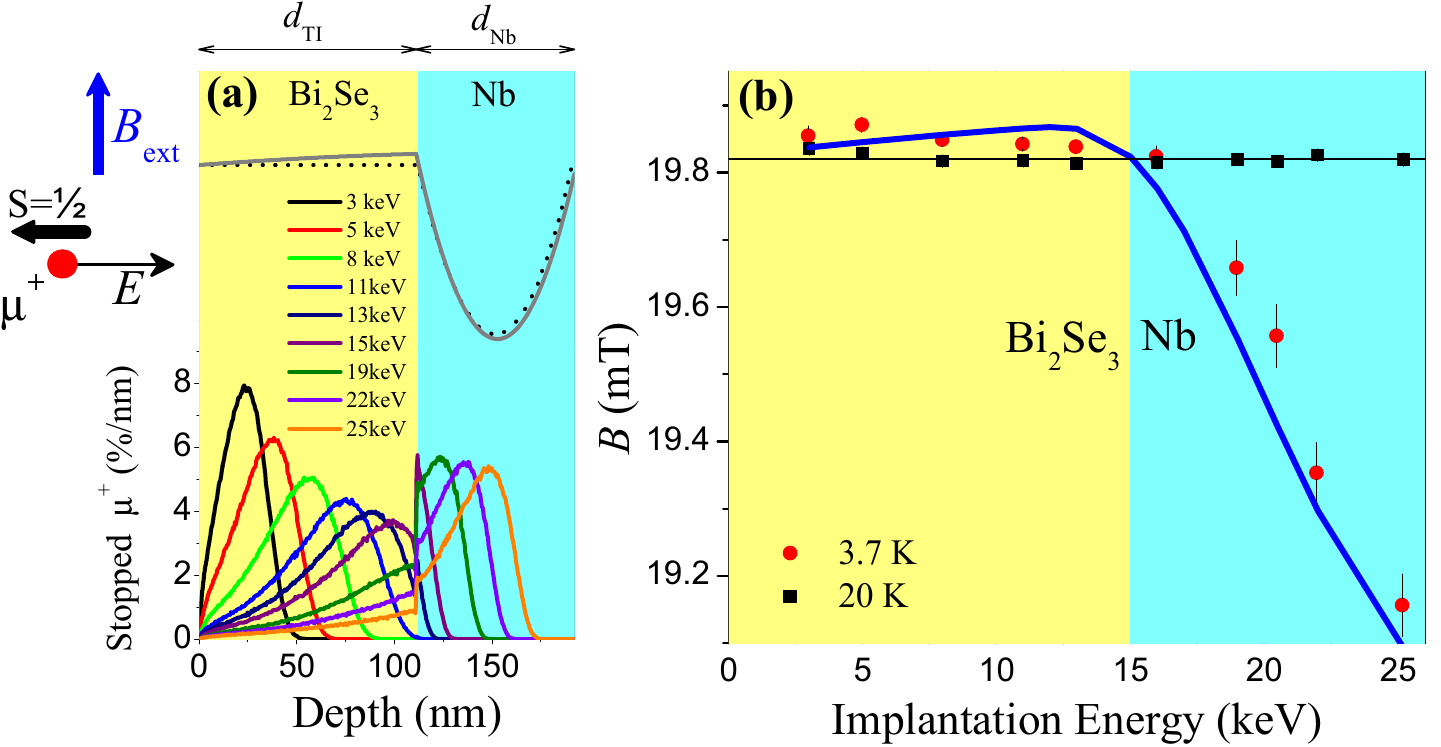}
  \caption{ \textbf{(a)} (top) Schematic showing the \mSR\ geometry
    where the applied field is parallel to the surface and
    perpendicular to the muon spin and momentum directions. The lines
    depict the expected magnetic field depth profile due to screening
    of the applied field inside the heterostructure with (solid gray)
    and without (dotted black) proximity induced superconductivity in
    the \BiSe\ layer.  (bottom) Calculated muon implantation profiles
    at different implantation energies using Trim.SP
    \cite{morenzoni2002}.  \textbf{(b)} Measured local mean field as a
    function of implantation energy above (black squares) and below
    (red circles) the superconducting transition temperature. The
    black horizontal line represents the applied field value,
    $B_\mathrm{ext}$, and the blue line is a fit to the theoretical
    model (see text).}
  \label{fig:depth}
\end{figure*}
One of the peculiarities of such a state is that the sign of the
Meissner screening can be reversed in some cases, causing an
attraction instead of a repulsion of external magnetic
fields. Paramagnetic Meissner response has been predicted in proximity
structures~\cite{Tanaka2005,Asano2011,Higashitani_prl2011,Mironov2012,Fominov_prb2015,LinderBalatsky2019PRM}
and in multiband SCs~\cite{Asano2015}. While such a paramagnetic
Meissner screening cannot be stable in the bulk, it has been observed
at interfaces, where the superconductivity is induced in a
nonsuperconducting layer by the proximity
effect~\cite{DiBernardo2015PRX}.  One particular \scg\ interface that
has attracted significant attention in recent years is between a
conventional $s$-wave SC and a 3D topological insulator
(TI). Fu~and~Kane have predicted that induced superconductivity in the
topological surface state (TSS) may have a $p_x+ip_y$ even-frequency
order parameter, that might allow for the stabilization of Majorana
bound states in vortex cores~\cite{Fu2008}. The latter are the key
ingredient in a proposal for fault-tolerant quantum
computation. Furthermore, the occurrence of Majorana zero modes can be
related to the presence of odd-frequency
\scg\ components~\cite{Asano2013,Stanev2014,Ebisu2015,Snelder2015}. Aside
from the conventional even-frequency superconductivity one should be
on a lookout for odd-frequency correlations induced at such
interfaces.

In this Letter, we provide experimental evidence of proximity induced
odd-frequency superconductivity in a heterostructure of \BiSe\ ($\sim
110$~nm) on Nb ($\sim 80$~nm). In particular, we measure the depth
dependence of the magnetic field parallel to the interface in the
Meissner state using low energy muon spin
rotation~\cite{Jackson2000,Suter2005PRB,Bakule2004,Yaouanc2011}
(LE-$\mu$SR) at the $\mu$E4 beam line~\cite{Prokscha2008} of the Swiss
Muon Source at PSI, Switzerland. This technique allows for a high
precision characterization of the magnetic field profile by measuring
the average Larmor precession frequency of the muons' spins as a
function of their implantation energy (and corresponding implantation
depth), as illustrated in Fig.~\refsubfig{fig:depth}{(a)}. Below the
superconducting transition temperature of Nb we observe conventional
diamagnetic Meissner screening in the Nb layer. In contrast, a
paramagnetic Meissner screening is observed in the \BiSe\ layer,
indicating a proximity-induced odd-frequency \scg\ component in the TI
layer. Our experimental results are complemented by theoretical
calculations supporting the appearance of an odd-frequency component
at the interface, extending deep into the TI.

\begin{figure}[ht]
\centering
\includegraphics[width=0.9\linewidth]{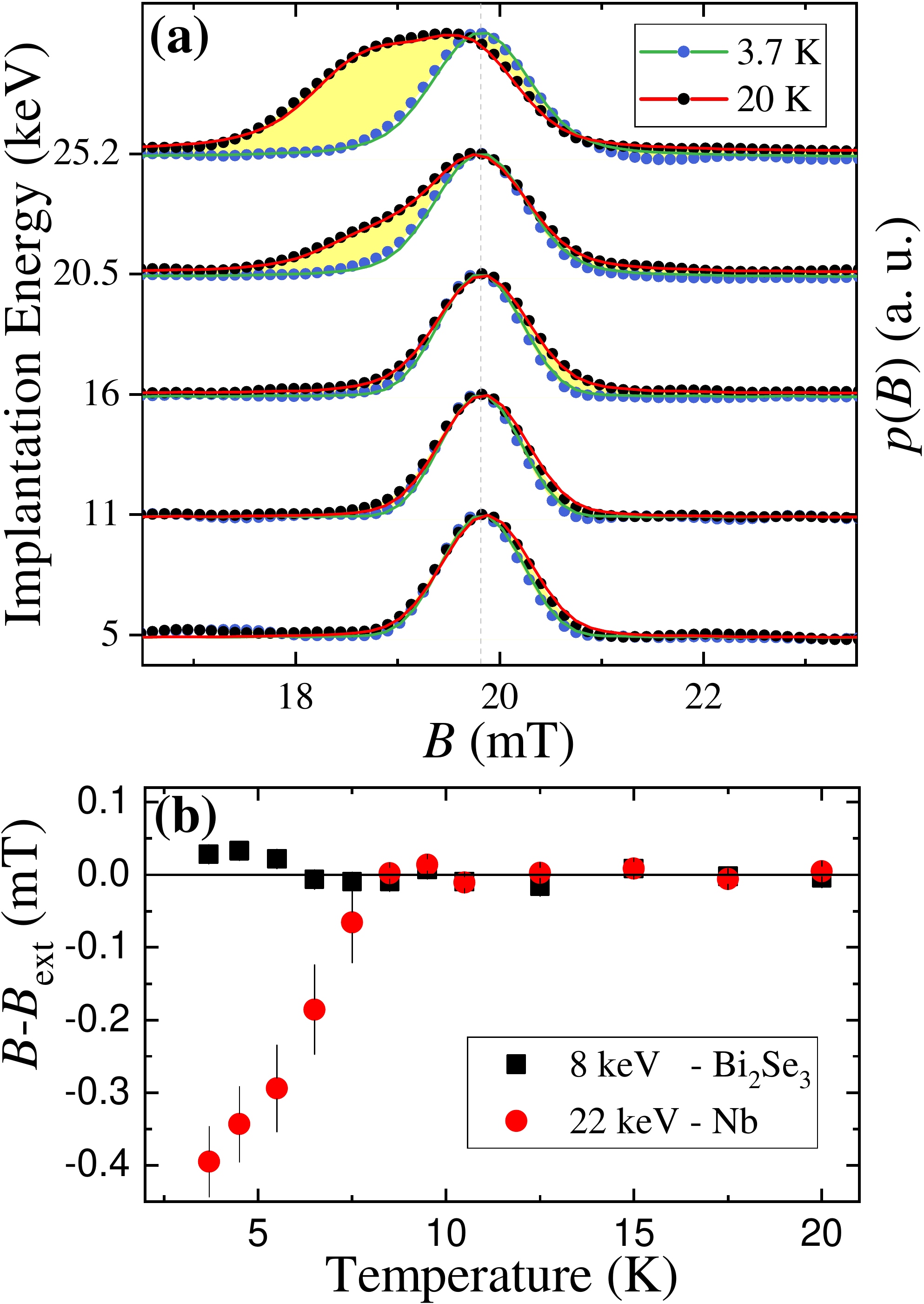}
\caption{ \textbf{(a)} Local field distribution calculated from fast
  Fourier transform of the muon polarization measured in the
  heterostructure, above and below the superconducting transition
  temperature. Different energies are offset for clarity.  The solid
  lines represent results from the fitting procedure (see
  SI\cite{SI}).  \textbf{(b)} Temperature dependence of the field
  shift in the \BiSe\ and Nb layers.}
\label{fig:temp}
\end{figure}

The local magnetic field as a function of implantation energy, $E$, is
shown in Fig.~\refsubfig{fig:depth}{(b)}. At a temperature of
\SI{20}{K}, i.e., in the normal state, the muons probe a
depth-independent magnetic field (black squares). At temperatures
below the superconducting transition of Nb, we observe a strong
variation in the measured field as a function of depth, featuring a
typical Meissner screening inside the Nb layer (red circles). In a
conventional metal-SC proximity structure, the induced superfluid
density in the metal results in a decreased mean field which increases
monotonically as a function of distance from the metal-SC interface,
reaching the applied field value far inside the
metal~\cite{Pambianchi94PRB,Flokstra2018PRL,Elvezio_private}. In
contrast, the behaviour in \BiSe/Nb is reversed; the field in the TI
is enhanced compared to the applied field value, $B_\mathrm{ext}$
(i.e., its value at \SI{20}{K}). This can be clearly seen in the fast
Fourier transform of the muon spin polarization spectra which
represent the local field distribution sensed by the muons
[Fig.~\refsubfig{fig:temp}{(a)}]. A careful inspection of the
temperature dependence of this effect shows that the paramagnetic
field shift in \BiSe\ occurs below the superconducting transition
temperature, $T_{\mathrm c} \approx 9$~K, of
Nb~[Fig.~\refsubfig{fig:temp}{(b)}].
This is in agreement with \Tc\ obtained from transport measurements
which show a sharp superconducting transition below 9~K (see the
Supplemental Material~\cite{SI}). We have also measured the magnetic
field as a function of depth for various $B_{\mathrm{ext}}$ values in
the range~\SI{7}{mT}~to~\SI{20}{mT}. We find that the local field at
low temperatures is proportional to the applied field, always
exhibiting a paramagnetic shift inside the TI. This indicates that the
induced superfluid density is almost field-independent within this
field range (see the Supplemental Material~\cite{SI}).

The observed paramagnetic shift of about \SI{0.2}{\percent} is too
large to be attributed to the demagnetization fields of the Nb layer,
which in a thin film are only relevant close to the
edges~\cite{SI}. Furthermore, a misalignment of the applied field with
respect to the interface could only reduce the mean field measured in
the sample. Another possible source for a positive shift are
microscopic demagnetization fields caused by the roughness of the
SC-TI interface. However, such stray fields should decrease
exponentially with distance away from the
interface~\cite{Tsymbal1994,Lindstrom14JEM}, which does not agree with
our observation. We can also exclude a temperature dependence of the
field in bulk \BiSe\ or any systematic deviations caused by the
experimental setup~\cite{Krieger2019}.

Odd-frequency components in superconductors are expected at any
ballistic interface \cite{Tanaka2007b} and in TIs, they may also occur
in the presence of in-plane gap gradients \cite{BlackSchaffer2012PRB},
exchange fields \cite{Yokoyama2012} and multigap odd-orbital coupling
\cite{BlackSchaffer2013PRB}. However, they are usually a subdominant
component compared to the even-frequency pairs. The total shielding
current given by ${\mathbf j}=-e^2(n_e-n_o){\mathbf A}/mc$, where
$n_e$ and $n_o$ are the superfluid densities of the even and odd
pairs, respectively, and $\mathbf A$ the vector-potential, should
therefore still be diamagnetic \cite{Mironov2012}. The observed
extension of the positive shift deep into the \BiSe\ layer, shown in
Fig.~\ref{fig:depth}(b), is a clear indication that not only the TSS
but also the bulk conduction band of \BiSe\ is relevant for the
proximity effect, as has also been pointed out by previous
studies~\cite{Xu2014}. Therefore, we conclude that the observed
paramagnetic Meissner screening is due to supercarriers induced into
the bulk conduction band of \BiSe\ by the proximity to
\scg\ Nb. Furthermore, this observation suggests that the
proximity-induced superconducting state is the unconventional
odd-frequency (Berezinskii) state. Such a nontrivial pairing state in
TIs has not been observed before.

To make a better comparison and illustrate the veracity of our
conclusion we provide a summary of theoretical calculations that
support our experimental results. We have developed a theoretical
description of proximity-induced superconductivity in the TI-SC
heterostructure based on a well-established two-orbital tight-binding
model for \BiSe~\cite{Rosenberg_Franz_prb2012,BlackSchaffer2013PRB}
(see the Supplemental Material~\cite{SI}). We show that the induced
odd-frequency pairing persists in the bulk of the TI, and that it
dominates the Meissner effect near the TI-SC interface. Based on the
insights from this microscopic model, we propose a theoretical depth
profile of the magnetic field, derived from Maxwell's equations and
linear response theory, that quantitatively fits our experimental
data.

The band inversion in \BiSe\ implies that the bulk conduction and
valence bands are formed by two orbitals with different parity,
originating from hybridized Se and Bi $p_z$ states. This orbital
degree of freedom in the TI allows for the generation of odd-frequency
pairing components, in addition to the even-frequency ones
~\cite{Triola_adp2020}. For a proximity-coupled TI to a singlet
$s$-wave SC, the symmetry allowed odd- (even-) frequency components
are odd(even) in the orbital index~\cite{BlackSchaffer2013PRB}. In
addition to the dominant $s$-wave singlet pairing, even- and
odd-frequency $p$-wave triplet components are present. Our
calculations show strong odd-frequency SC correlations that propagate
away from the interface due to the coupling between Nb and the bulk
electronic states of \BiSe. The two orbitals in the tight-binding
model correspond to top and bottom Se $p_z$ states in a quintuple
layer, each hybridized with the neighbouring Bi atoms
\cite{Fu2010PRL}. We assume stronger tunneling from the SC into the
orbital closest to the interface. This gives rise to odd-frequency
components that can be comparable or larger in magnitude than the
even-frequency ones over a wide range of frequencies. Details of this
effect and its depth dependence are given in the the Supplemental
Material \cite{SI}.

The penetration depth of the induced SC pairing in the TI layer
depends on the position of the chemical potential. When the chemical
potential is at the Dirac point of the TI, the induced SC pairing
amplitudes decay within approximately two quintuple layers from the
interface, i.e., a typical penetration depth of the TI surface
states. However, when the chemical potential is in the bulk conduction
band, as is the case for the (intrinsically) $n$-doped \BiSe\ sample
used here, we find finite SC pairing amplitudes far away from the
interface. We attribute this to coupling between the SC and {\em bulk}
TI states. This is consistent with the experimental observation of the
Meissner screening inside the TI layer shown in
Fig.~\ref{fig:depth}(b). The role of bulk states in the TI-SC
heterostructure has been previously pointed out for trivial
even-frequency pairing~\cite{Xu2014}. We extend this approach to allow
for odd-frequency pairing induced in the TI.

We will now focus on the interpretation of the positive field shift in
the observed Meissner screening.  In ordinary metals or
semiconductors, odd-frequency pairing leads to the paramagnetic
Meissner effect~\cite{Asano2015}. Recently, it has been pointed out
that in materials with Dirac dispersion, interband contributions to
the current-current correlation function can give rise to a
diamagnetic Meissner response \cite{Mizoguchi2015}. This has been
extended to a low-energy model of a doped 3D TI, where it was found
that odd-frequency pairing can give rise to a diamagnetic response due
to interband effects \cite{Schmidt2020PRB}. More generally, the total
magnetic response can be para- or diamagnetic depending on the balance
between intra- and interband processes \cite{Schmidt2020PRB}. This
complication, however, may be less relevant for the system considered
here since the model of a bulk TI does not rely on the Dirac
Hamiltonian. Therefore, we attribute the paramagnetic shift to the
odd-frequency components in the bulk TI. Based on these
considerations, we developed a phenomenological model of the Meissner
effect.

We consider a TI-SC heterostructure with the interface at $z=0$, SC
(Nb) extending from $-d_\mathrm{SC}$ to $0$ and the TI (\BiSe) from
$0$ to $d_\mathrm{TI}$.
Similarly to Ref.~\onlinecite{DiBernardo2015PRX}, we solve a
differential equation for the vector potential $A_x(z)\equiv{A(z)}$
\begin{eqnarray}\label{A(z)_diff}
 \frac{d^2{A(z)}}{dz^2}=K_{xx}(z)A(z),
\end{eqnarray} 
where $K_{xx}(z)$ is the current-current correlation function, or the
Meissner kernel, which determines the magnetic response of the system.
The local magnetic field is calculated as $B(z)=dA(z)/dz$. We use
matching boundary conditions for $A(z)$ and $B(z)$ at the TI-SC
interface and set $B(z)=B_\mathrm{ext}$ outside the heterostructure.
Furthermore, we propose the following form of the Meissner kernel in
the TI and SC layers
\begin{equation}\label{K_xx}
 K_{xx}(z)=\left\lbrace
 \begin{array}{ll}
 \frac{1}{\lambda^{2}_{\mathrm{SC}}} & -d_{\mathrm{SC}}<z<0\\
 -\frac{1}{\lambda^{2}_{\mathrm{TI}}}e^{-2z/z_0} & 0<z<d_\mathrm{TI},
\end{array}\right. 
\end{equation} 
where $\lambda_\mathrm{SC}$ is the London penetration depth of the SC
while $\lambda_\mathrm{TI}$ and $z_0$ are the characteristic length
scales in the TI layer. Thus, we assume a conventional diamagnetic
Meissner screening in the SC which leads to exponential suppression of
the local magnetic field extending into the SC layer. On the TI side,
the kernel describes the proximity-induced odd-frequency screening
which exponentially decays over a persistence length $z_0$ from the
TI-SC interface.


We solve Eq.~(\ref{A(z)_diff}) with the kernel $K_{xx}$ given in
Eq.~(\ref{K_xx}) and with boundary conditions at all three interfaces
to obtain the theoretical depth profile of the local magnetic field,
$B_\mathrm{loc}^\mathrm{theo}(z)$. To fit the experimental data, we
use $B_\mathrm{loc}^\mathrm{theo}(z)$ to calculate the muon precession
signal averaged over its stopping distribution at a given $E$ (see
details in the Supplemental Material~\cite{SI}).  The minimization of
the fit was performed for all $E$ values (at $T=3.7$~K)
simultaneously, using the same theoretical field profile. We also
fixed the thickness of the layers obtained from RBS and XRR
measurements, $d_\mathrm{TI}=111.1$~nm and $d_\mathrm{SC}=81$~nm,
together with the applied field obtained from the measurements at
$T=20$~K, $B_{\mathrm{ext}}=19.818(2)$~mT. The parameters extracted
from this global fit [gray line in Fig.~\ref{fig:depth}(a)] give
$\lambda_\mathrm{TI}=$\SI{1.62(4)}{\micro m} and
$\lambda_{SC}=117.8(6)$~nm. In this fit $z_0=$\SI{1}{ \micro m} is
chosen to reflect the fact that superconductivity is induced in the
\BiSe\ bulk conductance states above the TI gap and, therefore, has a
long decay range inside the TI. We checked that the specific choice of
$z_0$ does not affect the quality of the fit. However, the exact value
of of $\lambda_\mathrm{TI}$ systematically depends on the choice of
$z_0$ and varies between \SI{0.90(2)}{\micro m} and
\SI{1.74(5)}{\micro m} for $z_0$ within [$d_\mathrm{TI}$,$\infty$). We
  assume that the the penetration depth in Nb $lambda_\mathrm{SC}$ is
  not modified significantly by the interface. However, note that the
  obtained value is much larger than that in the clean limit of Nb
  [\SI{27(3)}{nm} \cite{Suter2005PRB}]. Finally, we calculated the
  mean local field as a function of $E$ by averaging
  $B_\mathrm{loc}^\mathrm{theo}(z)$ over the corresponding muons'
  stopping distribution. This is plotted in Fig.~\ref{fig:depth}(b)
  (blue line) and exhibits a very good agreement with the mean field
  obtained independently, confirming again the validity of the
  obtained $B_\mathrm{loc}^\mathrm{theo}(z)$.

Hence, the proposed theoretical field profile, based on the assumption
of a large odd-frequency pairing amplitude in the bulk of the TI
supported by the microscopic model, explains qualitatively and
quantitatively the observed paramagnetic Meissner shift. The measured
local field profile shows an almost constant paramagnetic shift in the
TI extending at least $30$~nm from the TI-SC interface and a
conventional screening on the \scg\ Nb side. The paramagnetic shift
decreases gradually towards the opposite TI surface. Deviations of the
experimental data from the predicted profile, for example a small
positive shift of the mean magnetic field above $B_\mathrm{ext}$ on
the SC side just below the interface may be due to the limited
accuracy in calculating the stopping profiles using Trim.SP
[Fig.~\ref{fig:depth}(a)]. Furthermore, some of the experimental
details that may affect the local field profile, such as interface
roughness and inhomogeneous thickness of the layers, are not included
in the theoretical modelling. However, we except this to have a
negligible effect, since the roughness is on an the length scale of
one quintuple layer ($\sim$1.4~nm) \cite{SI}. Therefore, the magnetic
field response is set by the macroscopic length scales, such as the
penetration depth. We also point out that both the induced dominant
odd-frequency superconductivity and the underlying Nb
superconductivity are of $s$-wave type. Therefore the effects of
atomic disorder at the interface are not likely to degrade the induced
odd-frequency state.

In conclusion, we observe an intrinsic paramagnetic Meissner shift in
proximity-induced superconductivity in a \BiSe/Nb heterostructure. We
attribute this effect to odd-frequency superconductivity which
persists up to tens of nanometers away from the \BiSe/Nb
interface. This finding, which is supported by our theoretical
calculations, is the first observation of a bulk induced odd-frequency
superconducting state in a TI. Our results demonstrate that the
experimental phenomenology of superconductivity at TI interfaces is
richer than previously thought, and it highlights the potential of
TI-SC heterostructures for realizing novel electronic
states. Odd-frequency \scg\ components may be of particular importance
for the theoretical description of TI and semiconductor-based Majorana
heterostructures which operate in a nonzero magnetic field.

The authors thank A.~M.~Black-Schaffer, R.~M.~Geilhufe and J.~Linder
for helpful discussions.  This work is based on experiments performed
at the Swiss Muon Source (S$\mu$S), Paul Scherrer Institute, Villigen,
Switzerland.  JAK and ZS are supported by the Swiss National Science
Foundation (SNF-Grant No.~200021\_165910). The work of AVB and AP is
supported by VILLUM FONDEN via the Centre of Excellence for Dirac
Materials (Grant No.11744), Knut and Alice Wallenberg Foundation
(Grant No. KAW 2018.0104) and the European Research Council
ERC-2018-SyG HERO.  T.H. acknowledges the John Fell Oxford University
Press (OUP) Research Fund, and thanks RCaH for their hospitality and
Liam J. Collins-McIntyre and Liam B. Duffy for help with the MBE
growth.

\bibliographystyle{apsrev}

\bibliography{Biblio}

\end{document}


\title {Supplementary Material\\Proximity-induced odd-frequency superconductivity in a topological insulator}

\author{Jonas~A.~Krieger}
\affiliation{Laboratory for Muon Spin Spectroscopy, Paul Scherrer 
Institute, CH-5232 Villigen PSI, CH}
\affiliation{Laboratorium f\"ur Festk\"orperphysik,  ETH Z\"urich, CH-8093
Z\"urich, Switzerland}
\affiliation{Swiss Light Source, Paul Scherrer Institute, 
CH-5232 Villigen PSI, Switzerland}
\author{Anna~Pertsova}
\affiliation{Nordita, Roslagstullsbacken 23, SE-106 91 Stockholm, Sweden}
\author{Sean~R.~Giblin}
\affiliation{School of Physics and Astronomy, Cardiff University, Cardiff, 
CF24 3AA, UK}
\author{Max D\"obeli}
\affiliation{Ion Beam Physics, ETH Z\"urich, Otto-Stern-Weg 5, CH-8093 Z\"urich, Switzerland}
\author{Thomas~Prokscha}
\affiliation{Laboratory for Muon Spin Spectroscopy, Paul Scherrer 
Institute, CH-5232 Villigen PSI, CH}
\author{Christof W. Schneider}
\affiliation{Laboratory for Multiscale Materials Experiments, Paul Scherrer Institut, CH-5232 Villigen PSI, Switzerland}
\author{Andreas~Suter}
\affiliation{Laboratory for Muon Spin Spectroscopy, Paul Scherrer 
Institute, CH-5232 Villigen PSI, CH}
\author{Thorsten~Hesjedal}
\affiliation{Department of Physics, Clarendon Laboratory, University of 
Oxford, Oxford, OX1 3PU, UK}
\author{Alexander~V.~Balatsky}
\email[Corresponding author: ]{avb@nordita.org}
\affiliation{Nordita, Roslagstullsbacken 23, SE-106 91 Stockholm, Sweden}
\affiliation{Department of Physics, University of Connecticut, Storrs, CT 06268, USA}
\author{Zaher~Salman}
\email[Corresponding author: ]{zaher.salman@psi.ch}
\affiliation{Laboratory for Muon Spin Spectroscopy, Paul Scherrer 
Institute, CH-5232 Villigen PSI, CH}

\date{\today}


%

\maketitle

\renewcommand{\theequation}{S\arabic{equation}}
\renewcommand{\thefigure}{S\the\value{figure}}
\renewcommand{\thetable}{S\the\value{table}}

\section{Experimental Methods}\label{sec:Exp}
The sample was prepared by sputtering a Nb layer on top of sapphire
substrate and transferred {\em ex situ} to a molecular beam epitaxy
(MBE) chamber (Createc GmbH). The Nb surface was then cleaned from any
oxide impurities, followed by MBE growth of a \BiSe\ layer on top of
Nb at 200\,$^\circ$C, and finally a cool-down under Se flow to
25\,$^\circ$C \cite{Collins-McIntyre2014AA}. A standard effusion cell
was used for the deposition of Bi (99.9999\% purity) while Se
(99.9999\% purity) was sublimated out of a cracker source (Createc
GmbH). A Bi:Se flux ratio of 1:10 was used to reduces the formation
\BiSe\ anti-site defects and Se vacancies.

Low energy muon spin spectroscopy measurements were performed on the
$\mu$E4 beamline of the Swiss Muon Source at \PSI\ in Villigen,
Switzerland~\cite{Prokscha2008}. In these measurements, fully
polarized muons are implanted at a given implantation energy, $E$, and
the evolution of their polarization is monitored via their anisotropic
beta decay. The probing depth inside the sample was tuned by varying
$E$ in the range \SI{3}{keV} to \SI{25}{keV}. The decay positrons were
detected by a set of detectors placed around the sample region. The
sample was cooled with a $^4$He flow cold finger cryostat. To suppress
the signal from muons missing the sample, the backing plate of the
sample was sputtered with a thin layer of Ni~\cite{Saadaoui2012PP}.  The
muon stopping profiles in the sample were modeled with the
\texttt{Trim.SP} code~\cite{morenzoni2002}.  As input to these
calculations we used the area density measured with Rutherford
backscattering (RBS) and the thicknesses determined with X-ray
reflectometry (XRR), as shown below.  The measured $\mu$SR asymmetry
(polarization) spectra were analyzed using the \texttt{Musrfit}
software~\cite{Suter2012PP}.  In a magnetic field applied perpendicular
to the initial muon spin direction, those time spectra can be
understood as the Fourier transform of the local field distribution
averaged over all muon stopping sites~\cite{Yaouanc2011}.

\section{\mSR{} Data Analysis}
\begin{figure}[h!]
  \includegraphics[width=.8\linewidth]{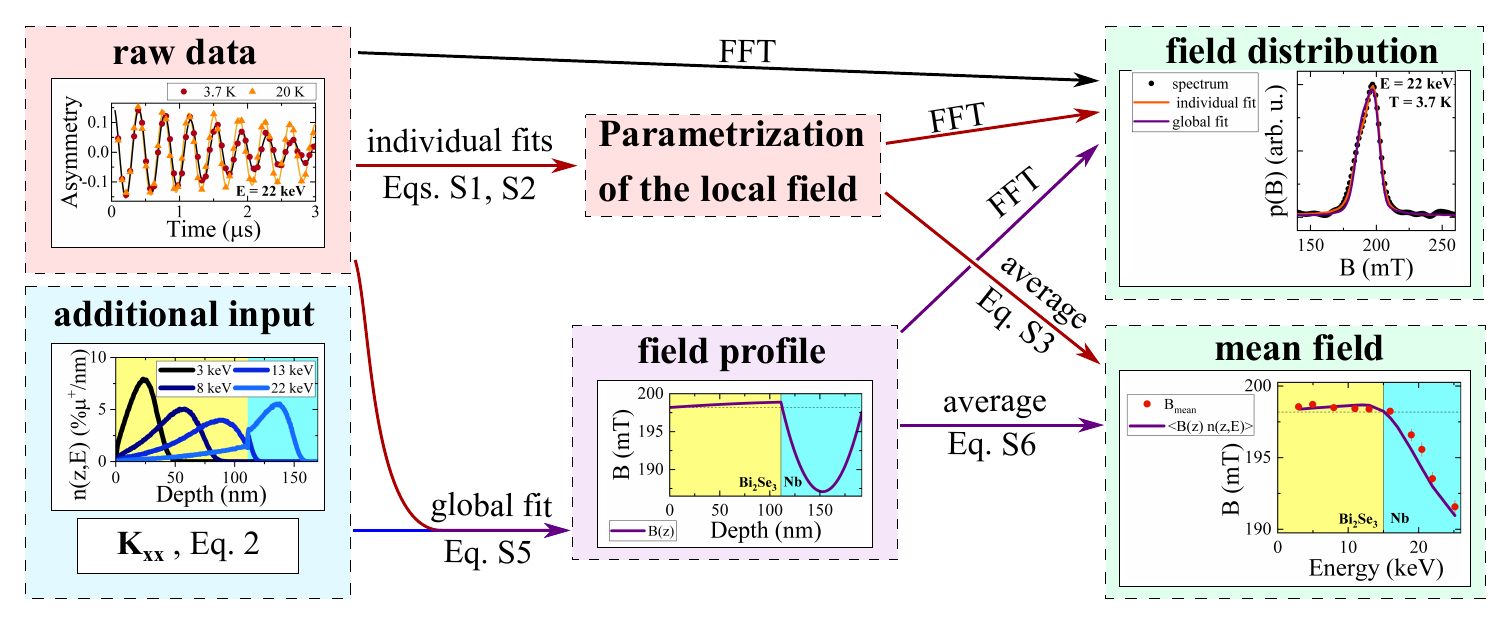}
  \caption{Schematic overview of the analysis procedures used on the
    experimental \mSR\ spectra.  }
  \label{fig:mSRanalysis}
\end{figure}

To confirm that the observed behavior is not an artifact of the
analysis procedure, we have analyzed the raw data using two
independent approaches. In the first approach, we parametrize each
individual spectrum separately to calculate the mean field sensed by
the muons at a given $E$ and temperature, $T$. In the second approach,
we use the theoretical field profile to model the muon spin
polarization for all values of $E$ and $T$ simultaneously, using the
corresponding stopping distributions. Both methods are described in
detail below and a schematic overview is given in
Fig.~\ref{fig:mSRanalysis}.

\subsection{Parametrization of individual \mSR\ spectra}
The asymmetry of the individual spectra can be fitted to an
exponentially damped oscillation,
\begin{equation}\label{eq:se}
 A(t)=A_0 e^{-\sigma t}\cos{(\gamma_\mu B t + \varphi)},
\end{equation}
where $A_0$ is the initial asymmetry, $\sigma$ is the depolarization
rate, $\gamma_\mu=2\pi\times\SI{135.5}{MHz/T}$ the muon's gyromagnetic
ratio and $\varphi$ is a detector-dependent geometrical phase factor.

The Nb nuclear magnetic moments are significantly larger and have
higher natural abundance than those of Bi and Se. This leads to a
different muon spin depolarization rates in the Nb and
\BiSe\ layers. This effect is of little importance above the
superconducting transition temperature where the measured asymmetry
can still be described by Eq.~(\ref{eq:se}) at all implantation
energies. However, in the superconducting state, Eq.~(\ref{eq:se})
fails to fit the data due to the strong depth dependence of the field
across the sample, in particular at $E>15$~keV, as can be clearly seen
in Fig.~2(a) in the main text. Therefore, we used a sum of two
exponentially damped signals to fit spectra measured with $E \ge
15$~keV and $T\le8$~K,
\begin{equation}\label{eq:2se}
 A(t)=A_{0,\mathrm{TI}} e^{-\sigma_{\mathrm{TI}} t}\cos{(\gamma_\mu
   B_{\mathrm{TI}} t + \varphi)}+ A_{0,\mathrm{Nb}}
 e^{-\sigma_{\mathrm{Nb}} t}\cos{(\gamma_\mu B_{\mathrm{Nb}} t +
   \varphi)}.
\end{equation}
Here the subscripts $\mathrm{TI}$ and $\mathrm{Nb}$ indicate which
layer the signal originates from. The mean field is then given by
\begin{equation}\label{eq:Bmean}
 B_{\mathrm{mean}}=\frac{A_{0,\mathrm{TI}}B_{\mathrm{TI}}
   +A_{0,\mathrm{Nb}}B_{\mathrm{Nb}}}{A_{0,\mathrm{TI}}
   +A_{0,\mathrm{Nb}}}.
\end{equation}
Note that the exact details of the fit do not affect the obtained
value of $B_{\mathrm{mean}}$ significantly as long as the fit mimics
the spectra accurately. The values ob $B_{\mathrm{mean}}$ as a
function of $E$ are plotted in Fig.~1(b) in the main text.

\subsection{Fit of the phenomenological field distribution}
Instead of parametrizing individual \mSR{} spectra, it is possible to
directly model them using a single depth profile of the magnetic field
$B(z)$, see e.g.~Ref.~\cite{Suter2005PRB,Suter2006PB}. This requires
knowledge of the muon stopping distribution $n(z,E)$, which is
obtained from the \texttt{Trim.SP} simulations. For this analysis we
have taken the experimentally determined layer thickness from RBS and
XRR (see below), $d_{\mathrm{TI}}=\SI{111.1}{nm}$ and
$d_{\mathrm{Nb}}=\SI{81.0}{nm}$.  At a given depth, $z$, we can write
the time evolution of the muon spin polarization as
\begin{equation}\label{eq:muPol}
 P(t,z)=\left\{\begin{matrix}
       e^{-\sigma_{\mathrm{TI}} t}\cos{(\gamma_\mu B(z) t + \varphi)}
       &\quad 0 \leq z\leq d_{\mathrm{TI}},\\
       e^{-\sigma_{\mathrm{Nb}} t}\cos{(\gamma_\mu B(z) t + \varphi)}
       &\quad d_{\mathrm{TI}}<z \leq d_{\mathrm{TI}}+d_{\mathrm{Nb}}.
      \end{matrix}\right.
\end{equation}
Therefore, the \mSR\ asymmetry spectrum at a given $E$ is calculated
by performing the ensemble average over $z$,
\begin{equation}\label{eq:globalFit}
 A(t,E)= A_0 \int_0^\infty\left. n(z,E)P(t,z) \right. \mathrm{d}z,
\end{equation}
where $A_0$ is the initial asymmetry. To fit the experimental data, we
evaluated this integral discreetly on a grid of \SI{0.5}{nm}.  At
\SI{20}{K} we assumed a constant field inside the sample, i.e.
$B(z)=B_{\mathrm{ext}}$. In the superconducting state, we assumed
$B(z)$ to be the solution to the phenomenological Meissner kernel from
Eq.~(2) in the main text, using the externally applied field
$B_{\mathrm{ext}}$ as a boundary condition. The details of the
functional form of $B(z)$ are given below in
Eqs.~(\ref{A(z)_diff2})~and~(\ref{A(z)2}). For consistency, we fixed
$B_{\mathrm{ext}}$, $\sigma_{\mathrm{TI}}$ and $\sigma_{\mathrm{Nb}}$
to their values obtained from the fits at \SI{20}{K}
(Table~\ref{tab:4K}). Furthermore, we assumed a persistence length
$z_0=\SI{ 1}{\micro m}$ in all fits, where $z_0$ describes the
propagation of the Nb electrons into the TI. This large $z_0$ value
corresponds to proximity effect with the chemical potential of Nb
lying within the bulk conduction band of Bi$_2$Se$_3$. On the other
hand, if the chemical potential were within the TI band gap, there
should only be proximity to the surface states and $z_0$ would be on
the scale of the lattice parameter.  Our fitting procedure gives
reasonable parameters only for large $z_0$ values, although the
quality of the fit is not affected by the specific choice of
$z_0$. The results from these global fits of all spectra
simultaneously at \SI{20}{K} and at \SI{3.7}{K} for different
$B_\mathrm{ext}$ are listed in Table~\ref{tab:4K}. The errors quoted
for these parameters reflects only the statistical errors. For
example, the specific choice of $z_0$ does not affect the quality of
the fit but changes the value of $\lambda_\mathrm{TI}$. $z_0$ between
$d_\mathrm{TI}$ and $\infty$ results in $\lambda_\mathrm{TI}$ between
\SI{0.90(2)}{\micro m}~and~\SI{1.74(5)}{\micro m}.
\begin{table}[h!]
  {\centering
\begin{tabular}{lll|ll}\hline\hline
\multicolumn{3}{c|}{\SI{20}{K}} & \multicolumn{2}{c}{\SI{3.7}{K}}
\\ \hline
\multicolumn{1}{c}{$B_{\mathrm{ext}}$~(mT)}
&\multicolumn{1}{c}{$\sigma_{\mathrm{TI}}$~($\mu$s$^{-1})$}
&\multicolumn{1}{c|}{ $\sigma_{\mathrm{Nb}}$~($\mu$s$^{-1})$}
&\multicolumn{1}{c}{ $\lambda_{\mathrm{Nb}}$~(nm)}
&\multicolumn{1}{c}{ $\lambda_{\mathrm{TI}}$~($\mu$m)}
\\ \hline\hline
$\quad$ 19.818(2) $\quad$ & $\quad$ 0.027(2) $\quad$  & $\quad$ 0.251(6) $\quad$  & $\quad$ 117.8(6) $\quad$  & $\quad$ 1.62(4) $\quad$ \\
$\quad$ 9.724(2) $\quad$   & $\quad$ 0.028(2) $\quad$  & $\quad$ 0.286(6) $\quad$  & $\quad$ 109.3(8) $\quad$  & $\quad$ 1.15(3) $\quad$ \\
$\quad$ 7.226(3) $\quad$   & $\quad$ 0.021(2) $\quad$  & $\quad$ 0.289(7) $\quad$  & $\quad$ 109(1) $\quad$    & $\quad$ 1.14(4)$\quad$ \\
\hline\hline
\end{tabular}}
\caption{Results and statistical errors from a simultaneous fit of
  Eq.~\ref{eq:globalFit} to the spectra measured at \SI{20}{K} (left)
  and $\sim$\SI{3.7}{K} (right) for different applied fields. The
  parameters from \SI{20}{K} fit without errors were assumed fixed for
  the fits of $\sim$\SI{3.7}{K}.}\label{tab:4K}
\end{table}

\begin{figure}[h!]
  \includegraphics[width=.95\linewidth]{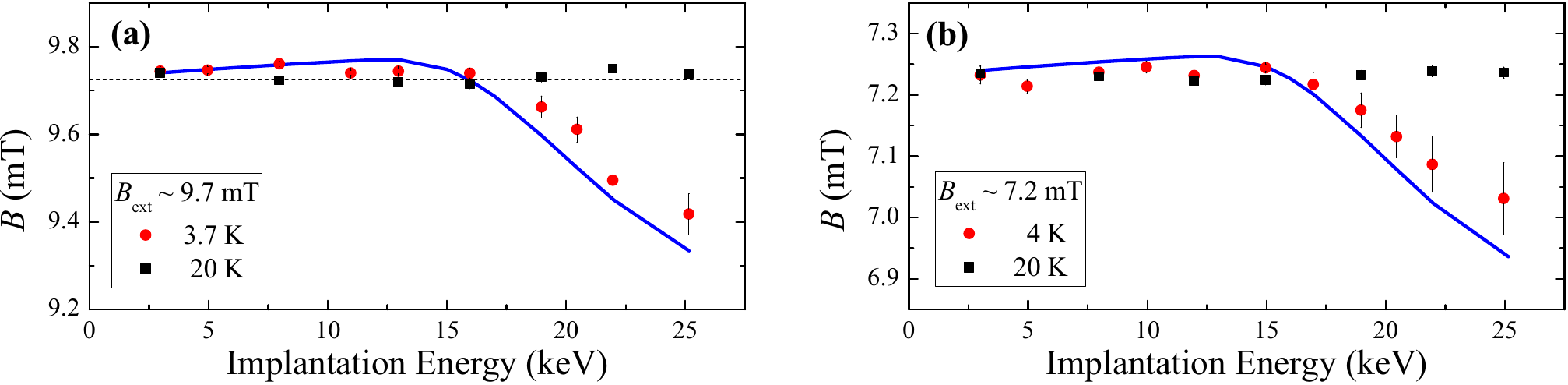}
  \caption{Measured local mean field as a function of implantation
    energy above (black squares) and below (red circles) the
    superconducting transition temperature. The black dashed line
    represents the applied field value, $B_\mathrm{ext}$, and the blue
    line results from the fit of the phenomenological field
    distribution.  }
  \label{fig:profile7_10mT}
\end{figure}

In order to compare the measured $B_\mathrm{mean}$
(Eq.~\ref{eq:Bmean}) with that calculated based $B(z)$ from the global
fit, we use,
\begin{equation}
 B_{\mathrm{mean}}^\mathrm{th}(E)=\left\langle B(z)
 n(z,E)\right\rangle=\int_0^\infty\left. B(z) n(z,E) \right.\mathrm{d}z .
\end{equation}
This is shown as solid lines in Fig.~\ref{fig:profile7_10mT} and
Fig.~1(b) of the main text.

\clearpage
\section{Additional Experimental Data}
\subsection{Transport Measurement}
\begin{figure}[h]
  \centerline{\includegraphics[width=0.4\linewidth]{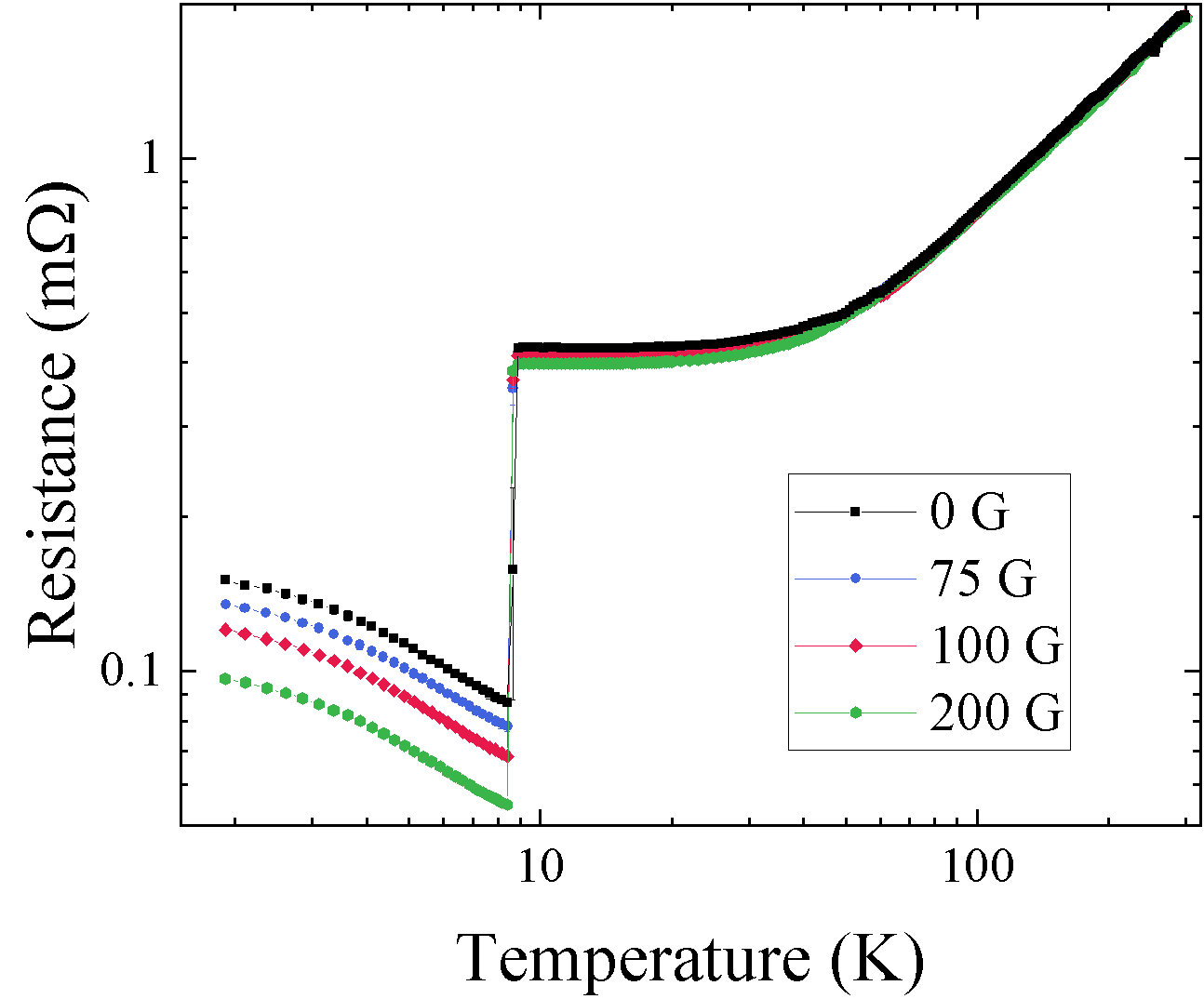}}
  \caption{Measured resistance as a function of temperature for
    different magnetic fields applied in-plane.}
  \label{fig:Res}
\end{figure}

The resistance of the sample as a function of temperature was measured
with pseudo van der Pauw configuration with four contacts placed close
to the edge of the film. In-plane magnetic fields of \SI{0}{mT},
\SI{7.5}{mT}, \SI{10}{mT}, and \SI{20}{mT} were applied during the
measurements. All curves show a drop in resistance below $T_{\mathrm
  c}\sim 10$~K, i.e. below the superconducting transition of Nb, see
Fig.~\ref{fig:Res}.
An upturn in the resistance below the superconducting transition of
Nb.  This is attributed to a residual resistance from the \BiSe\ layer
on top of Nb, where the measurement contacts are glued. Such an upturn
has been observed in other \BiSe\ samples and interpreted in terms of
quantum correction induced by electron-electron interactions
\cite{Wang2014SRa} or due to the presence of an impurity-band
\cite{Wiedmann2016PRB}.

\subsection{Rutherford backscattering}
\begin{figure}[h]
  \centering
  \includegraphics[width=0.4\linewidth]{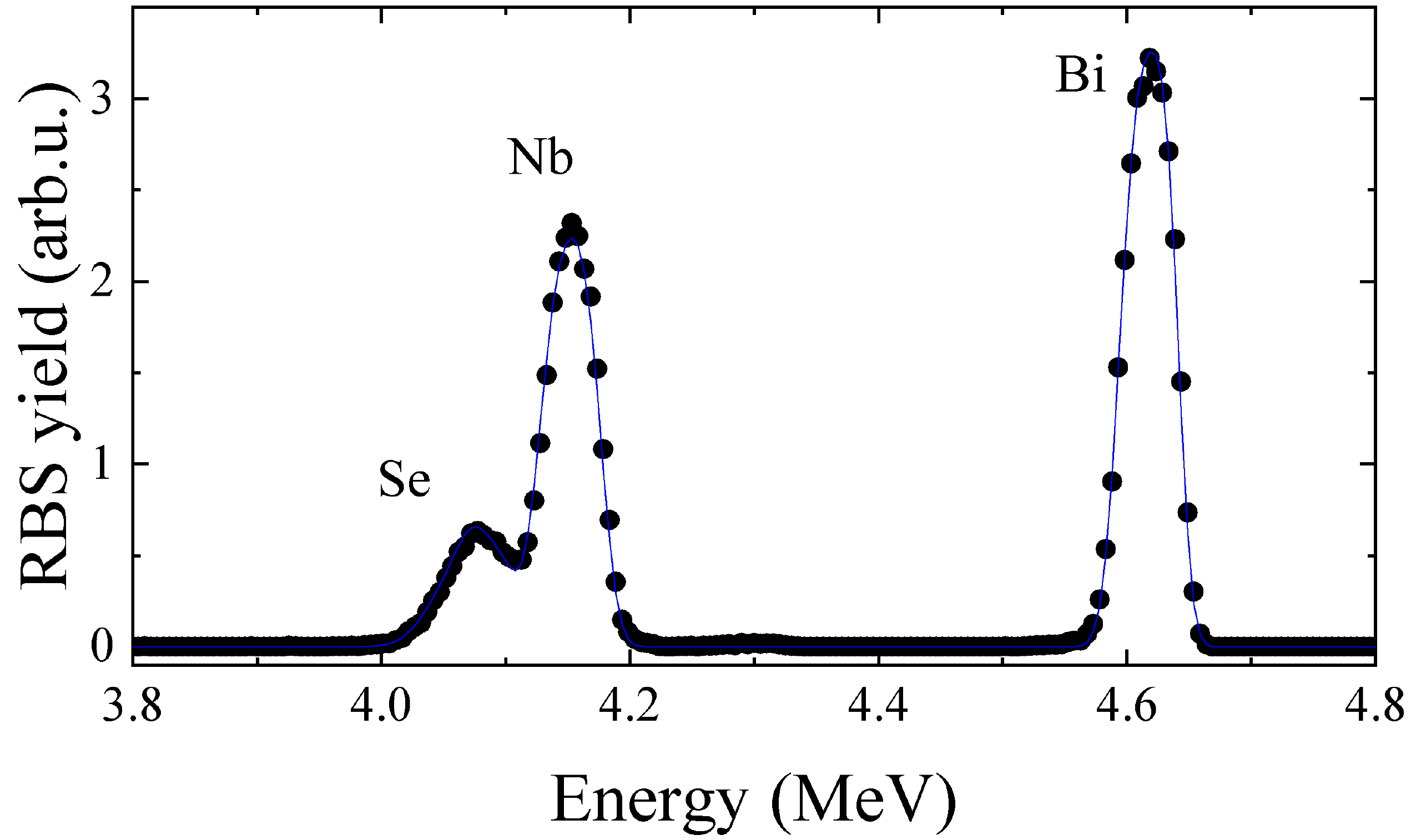}
  \caption{RBS spectrum of \SI{5}{MeV} He ions. The solid line shows
    the fitted curve.}
  \label{fig:RBS}
\end{figure}

The structure and stoichiometry of the sample was verified with RBS at
the Tandem-accelerator of ETH Zurich. Measurements were performed with
\SI{5}{MeV} $^4$He$^{2+}$ primary ions and a Si PIN diode detector
placed at \SI{168}{\degree}. Figure~\ref{fig:RBS} shows the RBS yield
as a function of the final He energy.
The spectrum was analyzed with the \texttt{RUMP}
software~\cite{Doolittle1985}. We found that the area densities of
Bi$_2$Se$_3$ and Nb are \SI{3.75(15)e17}{at.\per \centi\metre\squared}
and \SI{4.28(17)e17}{at.\per \centi\metre\squared}, respectively.

\subsection{X-ray reflectometry}
\begin{figure}[h]
  \centering
  \includegraphics[width=0.6\linewidth]{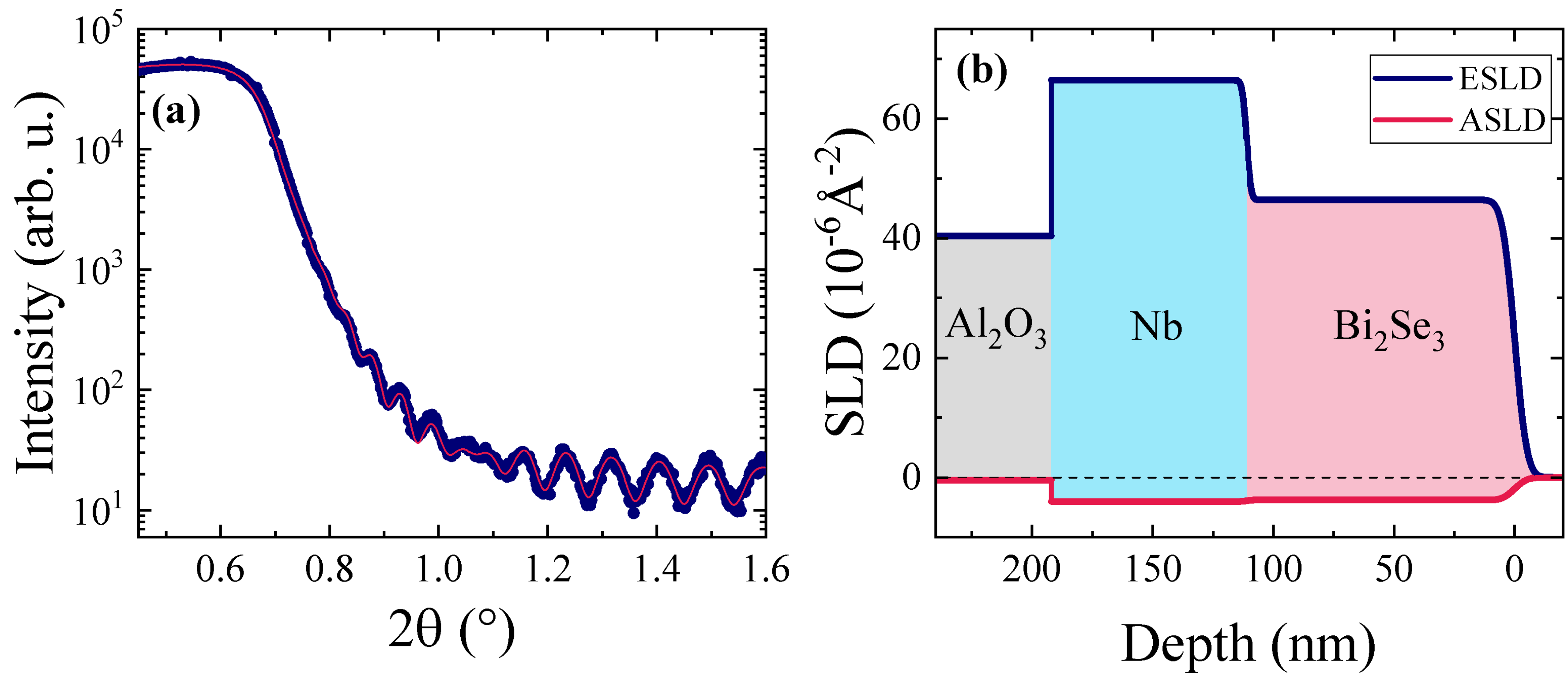}
  \caption{\textbf{(a)} X-ray reflectivity spectrum of the sample. The
    solid line corresponds to the elastic scattering length density
    (ESLD) and absorption length density (ASLD) shown in
    \textbf{(b)}.}
  \label{fig:XRR}
\end{figure}

To study the thickness of the layers we performed XRR on a Seifert
four-circle diffractometer, using the Cu-K$_\alpha$ line. The angle
dependence of the reflected intensity is shown in Fig.~\ref{fig:XRR}.
The corresponding scattering length densities were calculated with
\texttt{GenX}~\cite{Bjorck2007}. The resulting layer thicknesses are
$d_{\mathrm{TI}}=\SI{111.1}{nm}$ and $d_{\mathrm{Nb}}=\SI{81.0}{nm}$
with corresponding Bi$_2$Se$_3$/vacuum roughness of
$\sigma_{\mathrm{TI}}\approx\SI{4.0}{nm}$ and Bi$_2$Se$_3$/Nb
interface roughness of $\sigma_{\mathrm{Nb}}\approx\SI{1.4}{nm}$. The
Nb/Al$_2$O$_3$ substrate interface is almost perfectly flat.
Note that local stray fields caused by interface roughness at the
Bi$_2$Se$_3$/Nb interface should decay exponentially away from the
interface, on a length-scale of
$\sim$\SI{1}{nm}~\cite{Tsymbal1994}. This is an order of magnitude
shorter than the observed superconducting proximity.


\section{Theoretical modelling of a TI/SC heterostructure}
\subsection{Tight-binding model}
We consider a tight-binding model on a three-dimensional (3D) cubic
lattice which describes a 3D topological insulator (TI) coupled to a
superconducting layer (SC). The Hamiltonian of the system is written
as
\begin{equation}
 H=H_\mathrm{TI}+H_\mathrm{SC}+H_t,
\end{equation}
where $H_\mathrm{TI(SC)}$ is the Hamiltonian of the TI(SC) layer and
$H_t$ is the coupling Hamiltonian. To model the electronic structure
of Bi$_2$Se$_3$, we include two orbitals per atoms and spin-orbit
interaction~\cite{rosenberg_franz_prb2012,BlackSchaffer2013PRB,Fu2010PRL}. The
two orbitals represent two Se $p_z$-orbitals, originating from top and
bottom Se layers in a quintuple-layer unit cell of Bi$_2$Se$_3$, each
hybridized with its neighbouring Bi $p_z$-orbital. The hybridized Se
and Bi orbitals give rise to the inverted valence and conduction bands
of opposite parity. The conduction(valence) band has a predominant
contribution from Se(Bi) derived states. In the bulk, the two orbitals
can be transformed into each other by inversion with respect to the
central Se atom. However, in the TI/SC heterostructure the inversion
symmetry is broken and the two orbitals are shifted along the $z$-axis
with respect to the inversion center. In order to reflect this
asymmetry in the tight-binding model for a heterostructure, we assume
a larger tunnelling amplitude between the atoms of the SC layer and
one of the orbitals, corresponding the hybridized orbital closest to
the interface. The Hamiltonian of the TI layer can be written in the
spin and orbital basis as
\begin{equation}
 H_{\mathrm{TI}}=\sum_{\mathbf{k}}\Phi^{\dagger}_{\mathbf{k}}H_{\mathrm{TI}}(\mathbf{k})\Phi_{\mathbf{k}},
\end{equation}
where
$\Phi^{\dagger}=(a^{\uparrow\dagger}_{1\mathbf{k}},a^{\downarrow\dagger}_{1\mathbf{k}},a^{\uparrow\dagger}_{2\mathbf{k}},a^{\downarrow\dagger}_{2\mathbf{k}})$,
$a^{\sigma\dagger}_{\tau\mathbf{k}}$ ($a^{\sigma}_{\tau\mathbf{k}}$)
is the creation (annihilation) operator for an electron with spin
$\sigma$, orbital $\tau=1,2$ and momentum $\mathbf{k}$ inside the TI
layer.  The Hamiltonian matrix $H_{\mathrm{TI}}(\mathbf{k})$ is given
by,
\begin{equation}
 H_{\mathrm{TI}}(\mathbf{k})=\left(
 \begin{array}{lccr} 
  h_{11}(\mathbf{k}) & h_{12}(\mathbf{k}) & h_{13}(\mathbf{k}) & 0 \\
  h_{12}^{*}(\mathbf{k}) & h_{22}(\mathbf{k}) & 0 & h_{24}(\mathbf{k})  \\
  h_{13}^{*}(\mathbf{k}) & 0 & h_{33}(\mathbf{k}) & h_{34}(\mathbf{k})  \\
  0 & h_{24}^{*}(\mathbf{k}) & h_{34}^{*}(\mathbf{k}) & h_{44}(\mathbf{k}) 
 \end{array}
 \right),
\end{equation}
where the diagonal elements
$h_{\alpha\alpha}(\mathbf{k})=\gamma_0-\sum_{j}\gamma_i(e^{i{k_j}a}+e^{-i{k_j}a})$
($\alpha=1,...,4$, $j=x,y,z$) represent same-orbital (intra-orbital)
same-spin hopping, off-diagonal elements in spin index
$h_{12}(\mathbf{k})=-h_{34}(\mathbf{k})=-\lambda_x(e^{i{k_x}a}-e^{-i{k_x}a})+i\lambda_y(e^{i{k_y}a}-e^{-i{k_y}a})$
represent intra-orbital spin flips, and off-diagonal elements in
orbital index
$h_{13}(\mathbf{k})=h_{24}(\mathbf{k})=\epsilon-\sum_{j}t_j(e^{i{k_j}a}+e^{-i{k_j}a})+\lambda_z(e^{i{k_z}a}-e^{-i{k_z}a})$
represent inter-orbital same-spin hopping. Here, $a$ is the lattice
constant and $\mathbf{k}\equiv(\mathbf{k}^{||},k_z)$, with
$\mathbf{k}^{||}=(k_x,k_y)$ is the two-dimensional (2D) (in-plane)
momentum.

The Hamiltonian can be rewritten in concise form as,
\begin{equation}
    H_{\mathrm{TI}}=d_4\hat{I}+\sum_{\mu}{d_{\mu}}\Gamma_{\mu},
\end{equation}
where $\hat{I}$ is the identity matrix in spin and orbital space,
$d_4=\gamma_0-2\sum_{j}\gamma_j{\cos(k_{j}a)}$,
$d_0=\varepsilon-2\sum_{j}{t_j{\cos(k_j{a})}}$,
$d_j=-2\lambda_{j}\sin(k_j{a})$, $\Gamma_0=\tau_x\otimes{\sigma_0}$,
$\Gamma_x=-\tau_z\otimes{\sigma_y}$,
$\Gamma_y=\tau_z\otimes{\sigma_x}$, and
$\Gamma_z=\tau_y\otimes{\sigma_0}$, where $\tau$ and $\sigma$ are
Pauli matrices. The parameters of the model, $\gamma_0$, $\gamma_j$,
$t_j$, $\lambda_j$ and $\epsilon$ (with $j=x$, $y$,$z$) are fitted to
point~\cite{BlackSchaffer2013PRB}.  We use the following parameters:
$\gamma_0 = 0.3391$, $\gamma_{x,y} = 0.0506$, $\gamma_{z} = 0.0717$,
$\varepsilon= 1.6912$, $t_{x,y} = 0.3892$, $t_{z} = 0.2072$,
$\lambda_{x,y} = 0.2170$, and $\lambda_{z} = 0.1240$
eV~\cite{BlackSchaffer2013PRB}.  The eigenstates of $H_\mathrm{TI}$ can be
calculated analytically and are given by
$E_{\pm}=d_4\pm\sqrt{\sum_{\mu}d^{2}_{\mu}}$.

For the Nb layer, we consider a single layer of a singlet $s$-wave SC,
with the Hamiltonian given by
\begin{equation}
 H_{\mathrm{SC}}=\sum_{\mathbf{k},\sigma}\varepsilon_{\mathbf{k}}d^{\sigma\dagger}_{\mathbf{k},\sigma}d^{\sigma}_{\mathbf{k},\sigma}+
 \Delta_0\sum_{\mathbf{k}}(d^{\uparrow\dagger}_{\mathbf{k}}d^{\downarrow\dagger}_{-\mathbf{k}}
 +d^{\downarrow}_{-\mathbf{k}}d^{\uparrow}_{\mathbf{k}}),
\end{equation}
where $\mathbf{k}\equiv\mathbf{k}^{||}=(k_x,k_y)$ is the in-plane
momentum, $d^{\sigma\dagger}_{\mathbf{k}}$ ($d^{\sigma}_{\mathbf{k}}$)
is the creation (annihilation) operator for an electron with spin
$\sigma$ and momentum $\mathbf{k}$,
$\varepsilon_{\mathbf{k}}=-2[\cos(k_x{a}) +\cos(k_y{a})]$ is the
dispersion, and $\Delta_0$ is the superconducting order parameter in
the SC layer.

We assume nearest neighbors hopping between the atoms of the SC and
the TI interfacial layer, with different tunneling amplitudes for the
two orbitals of the TI atoms. The tunneling Hamiltonian is given by,
\begin{equation}
 H_{t}=-t_1\sum_{\mathbf{k},\sigma}d^{\sigma\dagger}_{\mathbf{k}}a^{\sigma}_{1\mathbf{k},i_z=0}
 -t_2\sum_{\mathbf{k},\sigma}d^{\sigma\dagger}_{\mathbf{k}}a^{\sigma}_{2\mathbf{k},i_z=0},
\end{equation}
where $t_{\tau}$, with $\tau=1,2$, are the orbital-dependent tunneling
matrix elements.  Here, the creation (annihilation) operators for TI
are labeled by the in-plane momentum $\mathbf{k}\equiv\mathbf{k^{||}}$
and the atomic index $i_z$, where $i_z=0$ for the interfacial
layer. As described above, in order to reflect the spatial asymmetry
between the two TI orbitals, we allow the tunnelling amplitudes $t_1$
and $t_2$ to be different (for the sake of definiteness, we assume
$t_1<t_2$).

\subsection{Proximity-induced Berezinskii paring at TI/SC interface}\label{odd-f}
Due to proximity to the SC layer, superconductivity is induced inside
the TI. Based on the symmetry analysis for an $s$-wave SC and a
multiband model of a TI with tetragonal
symmetry~\cite{BlackSchaffer2013PRB}, the possible induced pairing in the
TI are \textit{even-frequency} \textit{even-orbital}
and \textit{odd-frequency (Berezinskii)} \textit{odd-orbital}
spin-singlet and spin-triplet pairings.  In addition to the dominant
$s$-wave spin-singlet pairing, even- and odd-frequency $p$-wave
triplet components are present.  In order to find the pairing
amplitudes, we calculate the proximity-induced anomalous Green's
function at the interface. We define the electron, hole and anomalous
Green's functions as functions of momentum $\mathbf{k}$ and imaginary
time $t$,
\begin{eqnarray}
 \hat{G}^{\sigma\sigma}_{\tau\tau'}(\mathbf{k},t)&=&-\left\langle{\mathrm{T}_t
   a^{\sigma}_{\mathbf{k}\tau}}(t)a^{\sigma\dagger}_{\mathbf{k}\tau'}(0)\right\rangle
 \\ \hat{\bar{G}}^{\sigma\sigma}_{\tau\tau'}(\mathbf{k},t)&=&-\left\langle{\mathrm{T}_t
   a^{\sigma\dagger}_{-\mathbf{k}\tau}}(t)a^{\sigma}_{-\mathbf{k}\tau'}(0)\right\rangle
 \\ \hat{F}^{\sigma\sigma'}_{\tau\tau'}(\mathbf{k},t)&=&-\left\langle{\mathrm{T}_t
   a^{\sigma}_{\mathbf{k}\tau}}(t)a^{\sigma'}_{-\mathbf{k}\tau'}(0)\right\rangle
 \\ \hat{F}^{\sigma\sigma'\dagger}_{\tau\tau'}(\mathbf{k},t)&=&-\left\langle{\mathrm{T}_t
   a^{\sigma\dagger}_{-\mathbf{k}\tau}}(t)a^{\sigma'\dagger}_{\mathbf{k}\tau'}(0)\right\rangle.
\end{eqnarray} 
From the spin- and orbital-resolved anomalous Green's function, we
present the singlet and triplet pairing amplitudes that are either odd
or even in the orbital index. We introduce the following
singlet/triplet odd-frequency odd-orbital
($\mathcal{F}_{s/t}^{\tau\tau'(o)}$) and even-frequency even-orbital
($\mathcal{F}_{s/t}^{\tau\tau'(e)}$) pairing amplitudes,
\begin{eqnarray}
  \mathcal{F}_{s/t}^{\tau\tau'(o)}(\omega)&=&\frac{1}{\mathcal{N}}\sum_{\sigma\sigma'}\sum_{\mathbf{k}}S^{\sigma\sigma'}_{s/t}(\mathbf{k})[F_{\tau\tau'}^{\sigma\sigma'}(\mathbf{k},\omega)-F_{\tau'\tau}^{\sigma\sigma'}(\mathbf{k},\omega)]\label{eq:F_def1}
  \\ \mathcal{F}_{s/t}^{\tau\tau'(e)}(\omega)&=&\frac{1}{\mathcal{N}}\sum_{\sigma\sigma'}\sum_{\mathbf{k}}S^{\sigma\sigma'}_{s/t}(\mathbf{k})[F_{\tau\tau'}^{\sigma\sigma'}(\mathbf{k},\omega)+F_{\tau'\tau}^{\sigma\sigma'}(\mathbf{k},\omega)],\label{eq:F_def2}
\end{eqnarray}
where $\tau,\tau'=1,2$ and $\hat{S}_{s/t}$ is the spin-dependent
symmetry factor.  Here, $\hat{S}_s(\mathbf{k})=i\sigma_y$ for the
singlet component and
$\hat{S}_t(\mathbf{k})=2i(\mathbf{d}\cdot\sigma)\sigma_y$ for the
triplet component, with $\mathbf{d}$ a vector in spin space which
specifies the symmetry of the triplet state. For instance,
$\mathbf{d}=(k_x,k_y,0)$ for the triplet state with $A_{1u}$ symmetry
and $\mathbf{d}=(k_y,-k_x,0)$ for a triplet state with $A_{2u}$
symmetry. $\mathcal{N}$ is the normalization factor, equal to the
number of $k$-points in the Brillouin zone. The only non-zero
odd-frequency component is $\mathcal{F}_{s/t}^{12(o)}$, i.e. the
\textit{inter}-orbital component which is odd in the orbital
index. Both even-frequency \textit{intra-} and \textit{inter-}orbital
components, $\mathcal{F}_{s/t}^{\tau\tau(e)}$ ($\tau=1,2$) and
$\mathcal{F}_{s/t}^{12(e)}$ respectively, are present. We find
numerically, that the even-frequency intra-orbital component is
dominant, therefore we compare $\mathcal{F}_{s/t}^{12(o)}$ to
$\mathcal{F}_{s/t}^{\tau\tau(e)}$ in order to estimate the magnitude
of the odd-frequency pairing.
  
To the leading order in the tunneling matrix elements, the anomalous
Green's function in the TI is given by,
\begin{equation}
 \hat{F}^{\mathrm{TI}}=\hat{G}_{0}^{TI}\hat{H}_{t}\hat{F}_{0}^{\mathrm{SC}}\hat{H}^{\dagger}_{t}\hat{\bar{G}}_{0}^{\mathrm{TI}},
\end{equation} 
where $\hat{G}_0^{\mathrm{TI}}$ is the Green's function of TI in the
normal state, i.e. decoupled from the SC layer, and
$\hat{F}_0^{\mathrm{SC}}$ is the anomalous Green's function of an
isolated SC layer.  Here all Green's functions are matrices in spin
($\sigma$), orbital ($\tau$) and atomic indexes
($i\equiv{i_x,i_y,i_z}$), e.g.
$\hat{F}^{\mathrm{TI}}\equiv[\hat{F^{TI}}]^{\sigma\sigma'}_{\tau\tau',{i}i'}$. For
the inter-orbital pairing amplitude, we get
\begin{eqnarray}
 \hat{F}^{\mathrm{TI}}_{12}\equiv\hat{F}^{\mathrm{TI}}_{\tau=1,\tau'=2}  &=&  
 [\hat{G}_{0}^{TI}]_{11}\hat{T}_{1}\hat{F}_{0}^{\mathrm{SC}}\hat{T}^{T}_{2}[\hat{\bar{G}}_{0}^{\mathrm{TI}}]_{22}+\dots \\ \nonumber
  &=&  t_1 t_2[(\hat{G}_0^{\mathrm{TI}})_{11}]_{i_z=0,i_z'=0}\hat{F}_0^{\mathrm{SC}}[(\hat{G}_0^{\mathrm{TI}})_{22}]_{i_z=0,i_z'=0}+\dots,
\end{eqnarray} 
where $\hat{T_{\tau}}|=t_{\tau}\hat{I}$ and $\hat{I}$ is the identity
matrix.  After performing the Fourier transformation, the
inter-orbital pairing amplitude at the interface becomes
\begin{eqnarray}\label{F_12}
 \hat{F}^{\mathrm{TI}}_{12}(i\omega_n; i,i_z=0;i,i_z=0) &=& \\ \nonumber
 &  \frac{1}{\mathcal{N}}&\sum_{\mathbf{k}^{||}}
 \left[\left\lbrace\sum_{k_z}(\hat{G}_{0}^{TI})_{11}(i\omega_n,\mathbf{k}^{||},k_z)\right\rbrace
 \hat{F}_{0}^{\mathrm{SC}}(i\omega_n,\mathbf{k}^{||})\left\lbrace\sum_{k_z}(\hat{\bar{G}}_{0}^{\mathrm{TI}})_{22}
 (i\omega_n,-\mathbf{k}^{||},-k_z)\right\rbrace\right]+\cdots ,
\end{eqnarray}
where we used shorthand notation for the in-plane atomic indices,
$i=(i_x,i_y)$.
%
%
%
Note that here $\hat{F}^{\mathrm{TI}}_{12}$ is a matrix in spin
basis. Other pairing amplitudes in the orbital basis are calculated in
a similar way.

\begin{figure}[ht!]
\centering
\includegraphics[width=0.98\linewidth,clip=true]{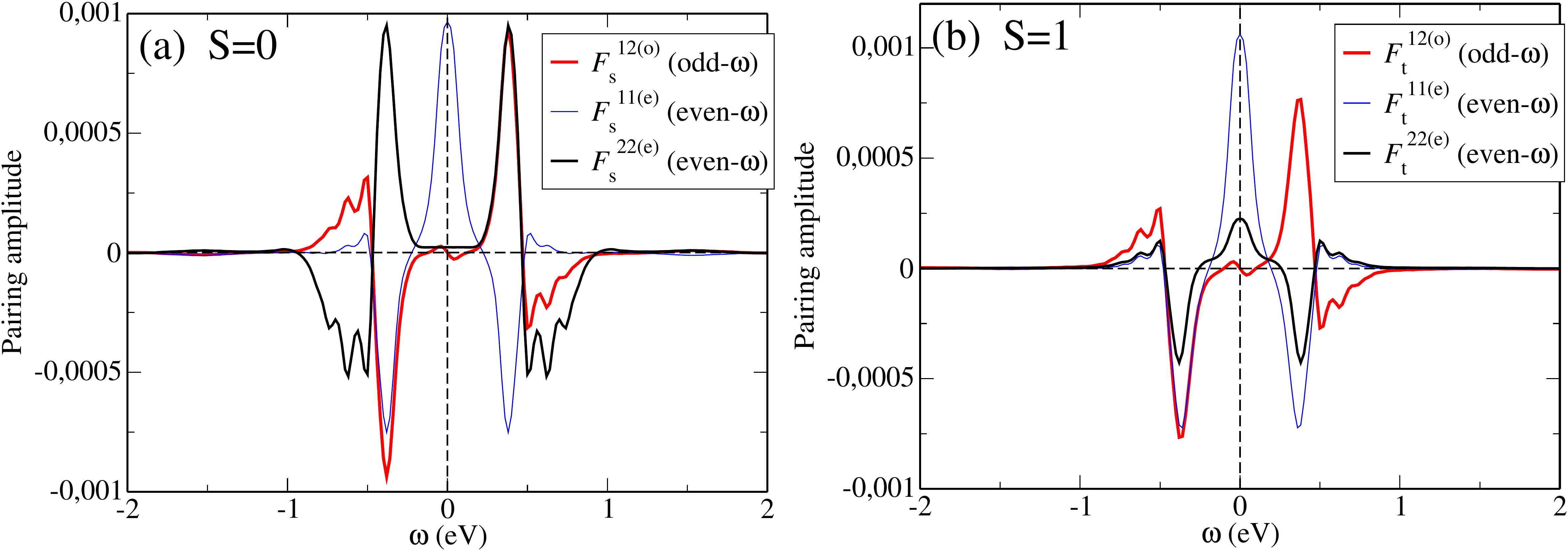}
\caption{The calculated odd- and even-frequency (a) spin-singlet (S=0)
  and (b) spin-triplet (S=1, A$_{2u}$ symmetry) pairing amplitudes
  (see Eqs.~(\ref{eq:F_def1})-(\ref{eq:F_def2}) for definitions) at
  TI/SC interface. Parameters are $\Delta_0=0.05$, $t_1/t_2=0.3$,
  $t_1=0.01$, $\mu_{\mathrm{SC}}=-3.0$. All energies are in eV.}
\label{fig1}
\end{figure}

\begin{figure}[ht!]
\centering
\includegraphics[width=0.6\linewidth,clip=true]{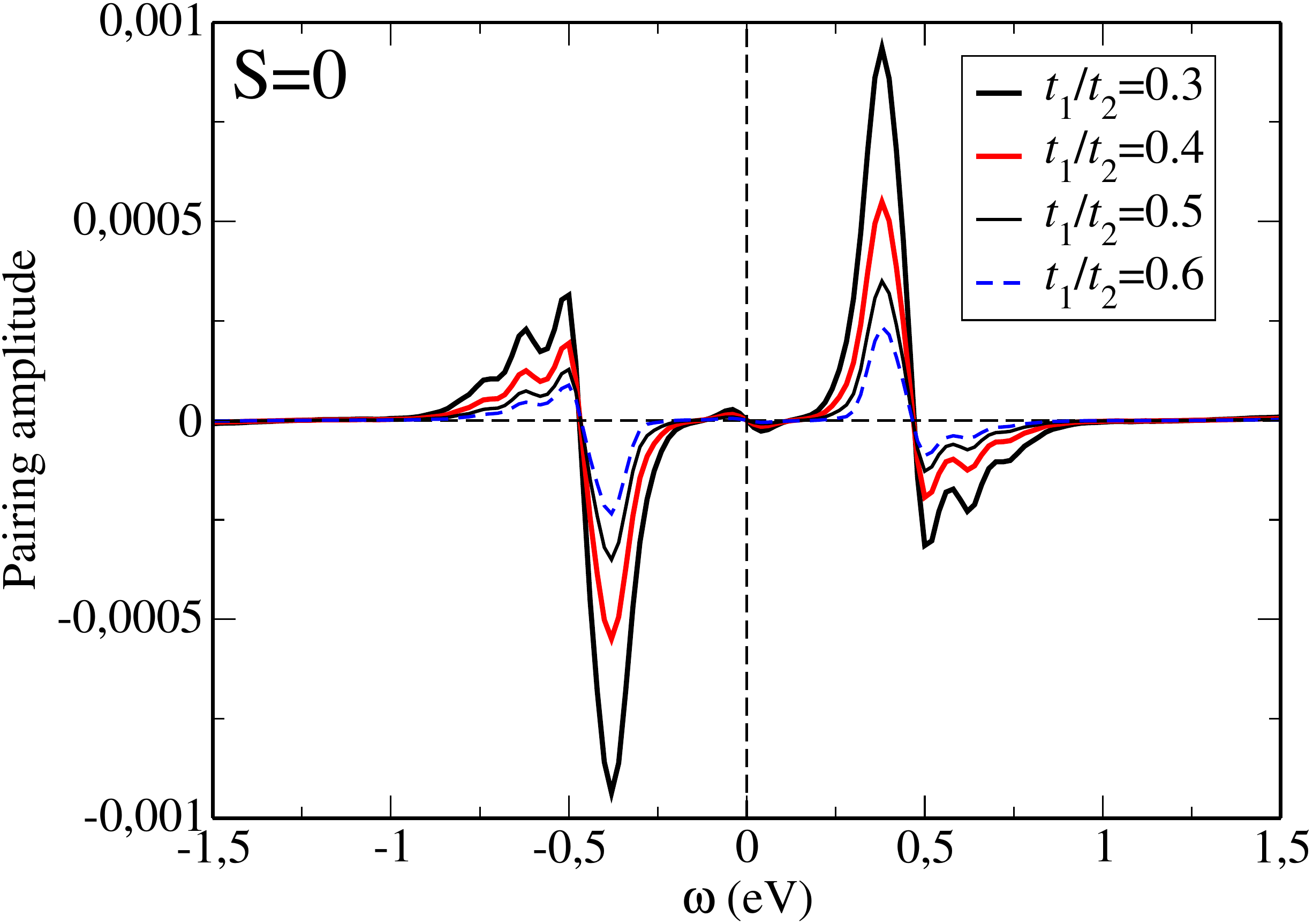}
\caption{The calculated odd-frequency spin-singlet (S=0) pairing
  amplitude at TI/SC interface for different tunneling ratios
  $t_1/t_2$.  Parameters are $\Delta_0=0.05$, $t_1=0.01$,
  $\mu_{\mathrm{SC}}=-3.0$. All energies are in eV.}
\label{fig1a}
\end{figure}

We also performed calculations using the tight-binding Hamiltonian of
a TI as described above, for a finite slab of TI coupled to a single
layer of SC. The induced pairing amplitudes calculated with this
approach are in qualitative agreement with the semi-analytical
approach based on the bulk model and second-order perturbation theory
[see Eq.~(\ref{F_12})]. However, the semi-analytical model is more
convenient for calculating the current-current correlation function
that determine the magnetic response of the system.  The odd- and
even-frequency pairing amplitudes at the interface (interfacial TI
layer), calculated using this approach, are shown in
Fig.~\ref{fig1}. The odd-frequency component is comparable to or
larger in magnitude than the even-frequency ones over a wide range of
frequencies. Figure~\ref{fig1a} shows the odd-frequency pairing
amplitude for different tunneling ratios $t_1/t_2$. The pairing
amplitude decreases with increasing $t_1/t_2$, i.e. with decreasing
the asymmetry in the coupling between the atoms of the SC and the two
TI orbitals, which is the source of the odd-frequency pairing.


\subsection{Enhancement of Berezinskii pairing due to coupling to bulk TI}\label{bulk-pairing}
\begin{figure}[ht!]
\centering
\includegraphics[width=0.98\linewidth,clip=true]{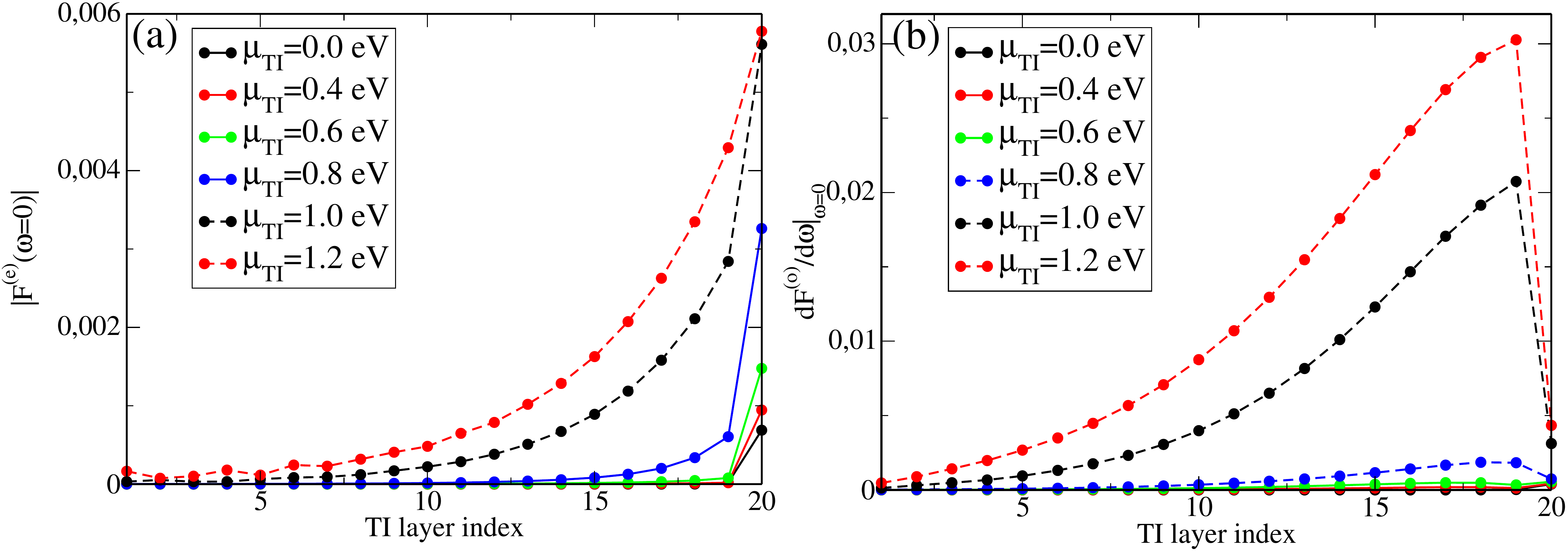}
\caption{The calculated (a) even-frequency and (b) odd-frequency
  spin-singlet (S=0) pairing amplitudes as a function of distance $z$
  from the interface for different values of the chemical potential
  inside the TI layer. The parameters used here are $\Delta_0=0.05$,
  $t_1/t_2=0.3$, $t_1=0.01$, $\mu_{\mathrm{SC}}=-2.0$.  All energies
  are in eV.  The maximum even-frequency amplitude is calculated as
  $\mathcal{F}^{e}(\omega=0)$ (here we show one of the intra-orbital
  amplitudes, $\mathcal{F}_{s}^{11(e)}$) and the magnitude of
  odd-frequency pairing is defined as
  $\frac{\partial{\mathcal{F}}^{(o)}(\omega)}{\partial\omega}|_{\omega=0}$.}
\label{fig_bulk_oddf}
\end{figure}

The depth dependence of the even- and odd-pairing amplitudes for
different positions of the chemical potential in the TI is presented
in Fig.~\ref{fig_bulk_oddf}.  We find that SC pairing extends deeper
into the bulk TI when $\mu_\mathrm{TI}$ is shifted away from the Dirac
point ($\mu_\mathrm{TI}=0$), i.e., into the conduction or valence
band. This is due to the fact that for a chemical potential larger
than the bulk TI bandgap, both the topological surface states and the
bulk TI states are present at the chemical potential and can therefore
couple to the SC.  This supports the experimental result that SC
exists not only at the interface where the TI surface states reside,
but also in the bulk with a large persistence length, $z_0$, such as
that used in our fits. This behavior has previously been seen for the
usual even-frequency pairing in the TI/SC heterostructures using
momentum-resolved photoemission spectroscopy~\cite{Xu2014}. Here, we
extend this result to odd-frequency pairing.  The details of the
interplay between the Dirac node and the Berezinskii state will be
explored more in a separate paper~\cite{Pertsova}.  Here we quote our
finding that the Berezinskii order-parameter defined as
$\Delta_\mathrm{Ber}\equiv\frac{\partial{\mathcal{F}}^{(o)}(\omega)}{\partial\omega}|_{\omega=0}$
grows into the bulk TI states as the doping pushes the chemical
potential away from the node.

\subsection{Current-current correlation function at the TI/SC interface} \label{kernel:calc}
To support the phenomenological model of the paramagnetic Meissner
effect in the TI/SC heterostructure, we calculate the current-current
correlation function from linear-response theory. In the case of
static and uniform magnetic field, the Meissner kernel,
$K_{xx}(\omega,q)$, is given by
\begin{eqnarray}\label{Kxx}
  K_{xx}(0,0)&=&T\sum_{\mathbf{k},\omega_n}\mathrm{Tr}[\hat{G}^{\mathrm{TI}}(\mathbf{k},\omega_n)\hat{j}_x(\mathbf{k})\hat{G}^{\mathrm{TI}}
  (\mathbf{k},\omega_n)\hat{j}_x(\mathbf{k})\\ \nonumber
  &+&\hat{F}^{\mathrm{TI}}(\mathbf{k},\omega_n)\hat{j}_x(\mathbf{k})\hat{F}^{\mathrm{TI}\dagger}(\mathbf{k},\omega_n)\hat{j}_x(\mathbf{k})], 
\end{eqnarray} 
where
$\hat{\mathbf{j}}(\mathbf{k})=\frac{\partial{\hat{H}^{TI}}(\mathbf{k})}{\partial{\mathbf{k}}}$. Here,
$\hat{j}_x$, $\hat{F}^\mathrm{TI}$ and $\hat{G}^\mathrm{TI}$ are
matrices in spin and orbital basis.  The Green's function for the TI
in proximity to the SC layer is calculated using second order
perturbation theory as explained in the previous section.

For the calculation based on the bulk model of the TI, $K_{xx}(0,0)$
gives the current-current correlation function at the interface and is
a constant dependent on the model parameters. Taking the corresponding
parameters for the TI Bi$_2$Se$_3$, our preliminary calculations based
on the bulk TI model show $K_{xx}(0,0)=-\kappa^2<0$, which is
consistent with the possible paramagnetic Meissner effect. Further
calculations for the full tight-binding model, as well as the role of
inter-band contributions to the Meissner effect, will be addressed
elsewhere.

%

\section{Theoretical parametrization of the local magnetic field profile}
We consider the TI/SC heterostructure with interface at $z=0$, Nb
extending from $-d_\mathrm{SC}$ to $0$ and TI from $0$ to
$d_\mathrm{TI}$.  The external magnetic field $\mathbf{B}_{ext}$ is
applied along the $y$-axis, $\mathbf{B}_{ext}=(0,B_{ext},0)$; $z$-axis
is the TI quintuple layer growth direction. For a phenomenological
description of the depth profile of the local magnetic field across
the TI/SC heterostructure, we use a combination of Maxwell's equations
and a linear response theory~\cite{DiBernardo2015PRX}.

From Maxwell's equations in transverse gauge, namely
$\nabla^2{\mathbf{A}}=-\mathbf{J}$,
$\mathbf{B}=\nabla\times\mathbf{A}$ for $\nabla\cdot{\mathbf{A}}=0$,
where $\mathbf{A}$ is the vector potential, $\mathbf{B}$ is the local
magnetic field and $\mathbf{J}$ is the current density, we obtain the
equation for the vector potential
\begin{eqnarray}\label{A(z)}
 \frac{d^2{A_x(z)}}{dz^2}=-J_x(z).
\end{eqnarray} 
From linear response theory, the vector potential and the current
density are related by
\begin{eqnarray}\label{linear_response}
J_x(z)=-K_{xx}(z)A_x(z),
\end{eqnarray} 
where $K_{xx}(z)$ is the current-current correlation function, or the
Meissner kernel. Substituting Eq.~(\ref{linear_response}) into
Eq.~(\ref{A(z)}), we get
\begin{eqnarray}\label{A(z)_diff}
 \frac{d^2{A_x(z)}}{dz^2}=K_{xx}(z)A_x(z).
\end{eqnarray} 
The local magnetic field inside the TI is given by $B(z)=B_{ext}+M(z)$
and from Maxwell's equations $B(z)=dA_x(z)/dz$.

We consider the following boundary conditions at the interfaces
\begin{eqnarray}\label{BC}
 B(d_\mathrm{TI})=B(-d_\mathrm{SC})=B_\mathrm{ext}\\
 B(0_{+})=B(0_{-})\equiv B_\mathrm{int}\\
 A(0_{+})=A(0_{-})\equiv A_\mathrm{int}.
\end{eqnarray} 

\subsection{Sign of the Meissner effect}
For now, we will neglect the depth dependence of the current-current
correlation function and assume that $K_{xx}(z)=$const.  For
even-frequency superconducting pairing, $K_{xx}(z)=\kappa^2>0$, which
gives a diamagnetic Meissner effect. The solution of
Eq.~(\ref{A(z)_diff}) gives,
\begin{eqnarray}\label{Meissner_even}
 M^{e}(z)=\frac{\cosh(\kappa{z})}{\cosh(\kappa)}-1.
\end{eqnarray} 
Here we have normalized the $z$-coordinate to the TI layer thickness,
$z\rightarrow{z/d_\mathrm{TI}}$.

It can be shown analytically for systems with quadratic dispersion
that for any odd-frequency superconducting pairing, regardless of its
origin, $K_{xx}(z)=-\kappa^2<0$~\cite{Asano2015}.  The magnetization
is given by,
\begin{eqnarray}\label{Meissner_odd}
 M^{o}(z)=\frac{\cos(\kappa{z})}{\cos(\kappa)}-\kappa.
\end{eqnarray} 
Hence, the magnetization and, as a result, the induced magnetic field
$B(z)$ inside the TI are an oscillatory decaying function of the
distance. Therefore, $B(z)$ can be larger or smaller than
$B_\mathrm{ext}$ and the exact shape of the field profile is
determined by material parameters. However, in order to have a
paramagnetic shift, the sign of $K_{xx}$ must be opposite to that of
the conventional SC. As shown in the previous section, the calculation
based on microscopic tight-binding model and linear response theory
yields $K_{xx}(z=0)<0$ at the interface.  This further supports the
dominant odd-frequency pairing assumed in the phenomenological model.

It should be mentioned that recent theoretical studies on
odd-frequency pairing in Dirac materials showed that due to inter-band
terms in the Dirac Hamiltonian, both even- and odd-frequency
components can contribute to diamagnetic and paramagnetic Meissner
effect~\cite{Mizoguchi2015,Schmidt2020PRB}. This means that the
paramagnetic Meissner effect in Dirac materials is not necessarily due
solely to the odd-frequency components. However, in the system studied
here, the induced SC pairing has bulk nature due to the electron
doping in the sample which shifts the chemical potential away from the
Dirac node and into the bulk TI conduction band.  Therefore, we
attribute the paramagnetic Meissner shift to the odd-frequency
pairing.

For the TI/SC heterostructure, the induced superconducting pairing has
both the even and odd-frequency components, as shown in the
theoretical calculations. In order to have the overall paramagnetic
response, the odd-frequency component should be dominant. In addition,
the oscillating odd-frequency component should produce a positive
shift of the local magnetic field with respect to
$B_\mathrm{ext}$. Below we construct a theoretical parametrization of
magnetic field depth profile that describes well the experimental
data.

\subsection{Phenomenological Meissner kernel $K_{xx}$ in TI/SC heterostructure}\label{sec:PhenMeissner}
We assume exponential decay of the magnetic field inside the Nb layer,
from both interfaces (TI/SC and SC/vacuum or substrate).  The Meissner
kernel is given by
\begin{equation}\label{Kxx_Nb}
 K_{xx}(z)=-\frac{1}{\lambda^{2}_\mathrm{SC}},\qquad z<0
\end{equation} 
where $\lambda_\mathrm{SC}$ is the London penetration depth of the SC
layer.  This leads to the following solutions for the field and the
vector potential,
\begin{eqnarray}\label{A(z)_diff2}
 B(z)&=&c_1 e^{\frac{-z-d_\mathrm{SC}}{\lambda_\mathrm{SC}}}+c_2 e^{\frac{z}{\lambda_\mathrm{SC}}},\qquad z<0\\
 A_\mathrm{SC}(z)&=&-\lambda_\mathrm{SC}c_1 e^{\frac{-z-d_\mathrm{SC}}{\lambda_\mathrm{SC}}}+\lambda_\mathrm{SC}c_2 e^{\frac{z}{\lambda_\mathrm{SC}}},\qquad z<0
\end{eqnarray} 
The kernel inside the TI layer can be written as
\begin{eqnarray}\label{Kxx_TI}
 K_{xx}(z)=-a^2 e^{-2bz},\qquad z>0
\end{eqnarray} 
where $a\equiv 1/\lambda_\mathrm{TI}$ and $b\equiv 1/z_0$ are fit
parameters (see Eq.~(2) of the main text).  This leads to the
following solutions for the field and the vector potential
\begin{eqnarray}\label{A(z)2}
 B_\mathrm{TI}(z)&=&a e^{-bz} \left[ d_1 J_{1} \left(\frac{a}{b}e^{-bz}\right)+d_2 Y_{1}\left(\frac{a}{b}e^{-bz}\right)\right],\qquad z>0, \\
 A_\mathrm{TI}(z)&=&d_1 J_{0}\left(\frac{a}{b}e^{-bz}\right)+d_2 Y_{0}\left(\frac{a}{b}e^{-bz}\right),\qquad z>0
\end{eqnarray} 
where $J_{0,1}$ and $Y_{0,1}$ are $0^\mathrm{th}$ or $1^\mathrm{st}$
order Bessel functions of the first and second kind, respectively. The
values of $c_1$, $c_2$, $d_1$ and $d_2$ are determined by the boundary
conditions.  The results of the fit of the experimental field profile
using the Bessel-function based solution for the vector potential are
presented in Fig.~1 of the main text.

\section{Finite element modelling of TI/SC heterostructure}
Here we model the stray fields from a slab of thin superconducting
layer to rule out such effects as a source of the observed magnetic
field depth dependence in the TI/SC heterostructure. Although trivial,
we present these calculations for completeness.
\begin{figure}[htb]
  \centerline{\includegraphics[width=0.5\linewidth]{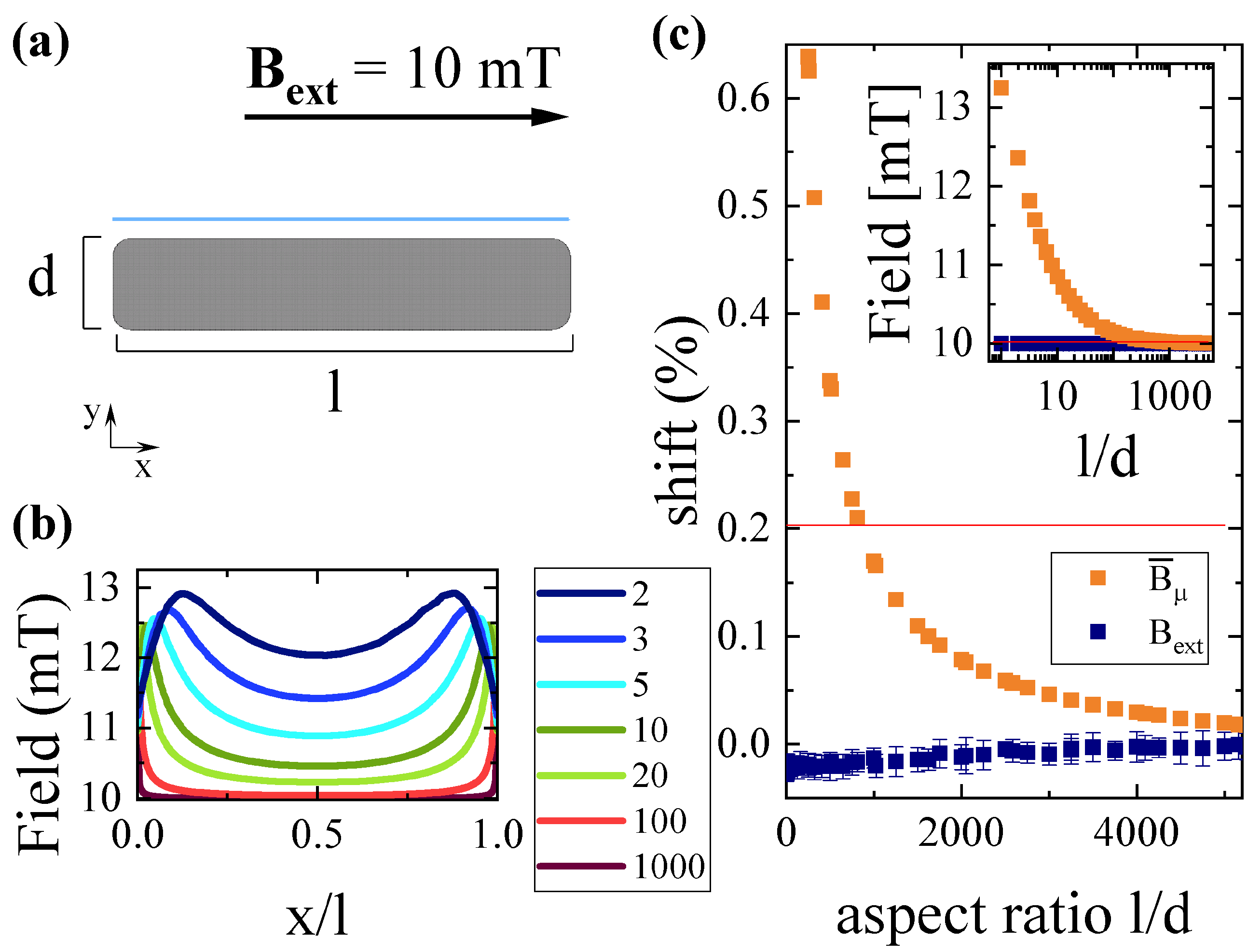}}
  \caption{Stray fields from a thin superconducting layer, modeled
    with 2d finite element analysis.  \textbf{(a)} Model geometry: A
    thin, ideal diamagnet with an aspect ratio of $l/d$ in an external
    field of \SI{10}{mT} was simulated.  The field was probed at the
    blue line $1/5$\,$d$ above the sample. The aspect ratio was varied
    by changing $l$.  \textbf{(b)} In-plane field along the blue line
    for different aspect ratios.  \textbf{(c)} Shift of the in-plane
    field with respect to \SI{10}{mT} as a function of aspect ratio
    (orange points). The blue points show the external field probed
    very far away from the sample. The inset depicts the same in a
    larger region and in absolute units. Experimentally
    $l/d\sim\SI{e5}{}$.}
  \label{fig:ARatio}
\end{figure}
\begin{figure}[htb]
  \centerline{\includegraphics[width=0.5\linewidth]{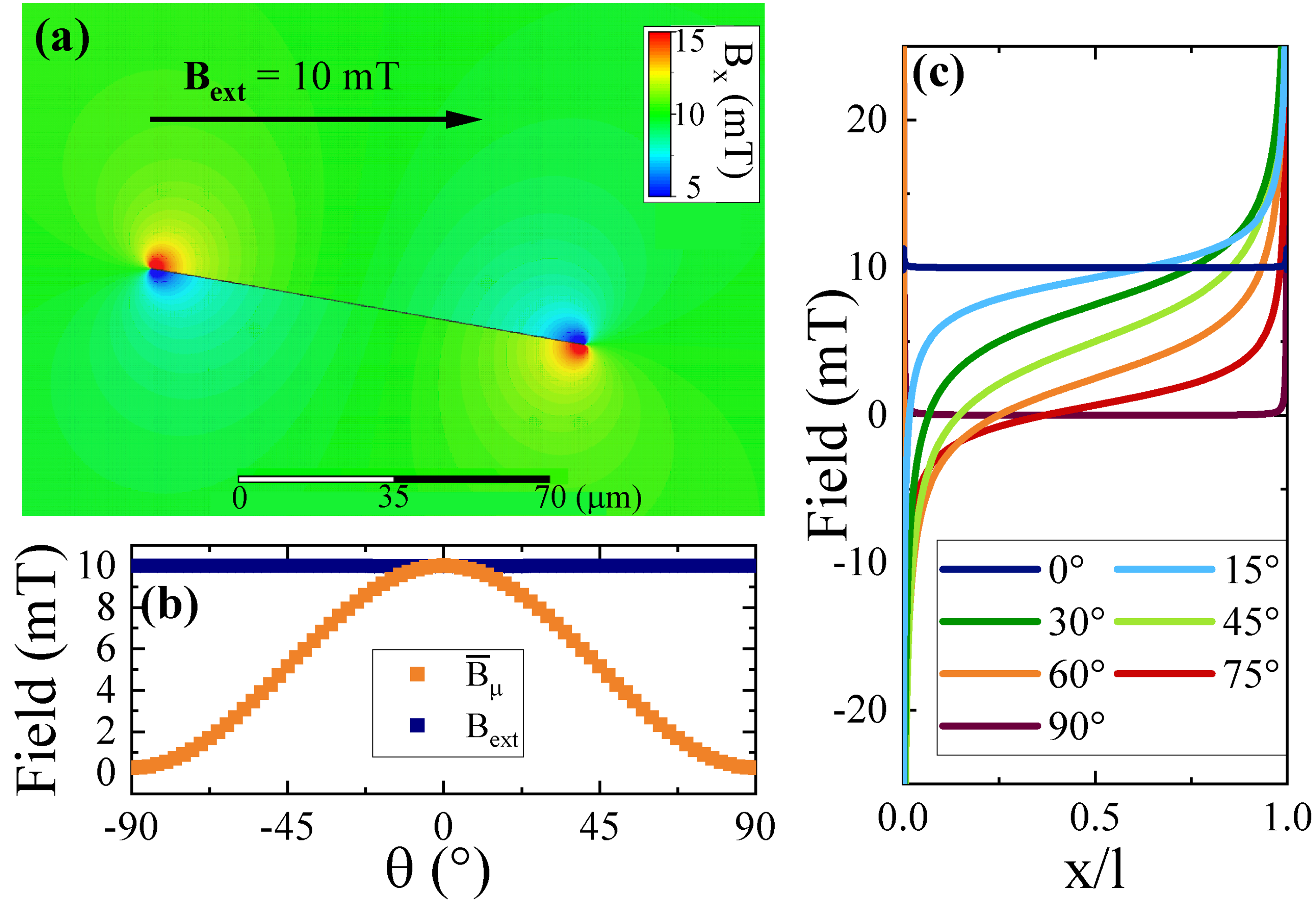}}
  \caption{Finite element analysis of a sample tilted with an angle
    $\theta$ with respect to the applied field.  \textbf{(a)} In-plane
    fields for $\theta=\SI{10}{\degree}$.  \textbf{(b)} Average
    in-plane field as a function of $\theta$ at the sample (orange
    points) and far away (blue points).  \textbf{(c)} In-plane field
    across the samples for different rotation angles $\theta$.}
  \label{fig:ARot}
\end{figure}

\clearpage
\bibliography{Biblio}